\documentclass{emulateapj}
\usepackage{apjfonts}
\usepackage{amsmath}
\usepackage{amssymb}
\usepackage[colorlinks=true,linkcolor=blue,citecolor=blue]{hyperref}

\newcommand{\kms}{\ensuremath{\mathrm{km~s}^{-1}}}
\newcommand{\msun}{\ensuremath{M_\odot}}
\newcommand{\Nifs}{\ensuremath{^{56}\mathrm{Ni}}}
\newcommand{\Cofs}{\ensuremath{^{56}\mathrm{Co}}}
\newcommand{\Fefs}{\ensuremath{^{56}\mathrm{Fe}}}

\newcommand{\beq}{\begin{equation}}
\newcommand{\eeq}{\end{equation}}
\newcommand{\beqar}{\begin{eqnarray}}
\newcommand{\eeqar}{\end{eqnarray}}
\newcommand{\rt}{r_{\rm t}}
\newcommand{\rs}{r_{\rm s}}
\newcommand{\Mbh}{M_{\rm bh}}
\newcommand{\rp}{r_{\rm p}}

\received{}
\accepted{}

\begin{document}
\bibliographystyle{apj}

\title{Optical Thermonuclear Transients From Tidal Compression of White Dwarfs as Tracers of the Low End of the Massive Black Hole Mass Function}

\author{Morgan MacLeod$^{1}$, James Guillochon$^{2,3}$, Enrico Ramirez-Ruiz$^{1}$, Daniel Kasen$^{4,5}$ and Stephan Rosswog$^{6}$ } 
\altaffiltext{1}{Department of Astronomy \& Astrophysics, University of California, Santa Cruz, CA 95064}
\altaffiltext{2}{Einstein Fellow}
\altaffiltext{3}{Harvard-Smithsonian Center for Astrophysics, The Institute for Theory and Computation, 60 Garden Street, Cambridge, MA 02138, USA}
\altaffiltext{4}{Department of Physics, University of California, Berkeley, CA 94720, USA.}
\altaffiltext{5}{Nuclear Science Division, Lawrence Berkeley National Laboratory, Berkeley, CA 94720, USA.}
\altaffiltext{6}{The Oskar Klein Centre, Department of Astronomy, AlbaNova, Stockholm University, SE-106 91 Stockholm, Sweden.}

\begin{abstract} 
In this paper, we model the observable signatures of tidal disruptions of white dwarf (WD) stars by massive black holes (MBHs) of moderate mass, $\approx 10^3 - 10^5 M_\odot$. When the WD passes deep enough within the MBH's tidal field, these signatures include thermonuclear transients from burning during maximum compression. We combine a hydrodynamic simulation that includes nuclear burning of the disruption of a $0.6 M_\odot$ C/O WD with a Monte Carlo radiative transfer calculation to synthesize the properties of a representative transient. The transient's emission emerges in the optical, with lightcurves and spectra reminiscent of type I SNe. The properties are strongly viewing-angle dependent, and key spectral signatures are $\approx 10,000$~\kms\ Doppler shifts due to the orbital motion of the unbound ejecta. 
Disruptions of He WDs likely produce large quantities of intermediate-mass elements, offering a possible production mechanism for Ca-rich transients.
Accompanying multiwavelength transients are fueled by accretion and arise from the nascent accretion disk and relativistic jet.  If MBHs of moderate mass exist with number densities similar to those of supermassive BHs, both high energy wide-field monitors and upcoming optical surveys should detect tens to hundreds of WD tidal disruptions per year. The current best strategy for their detection may therefore be deep optical follow up of high-energy transients of unusually-long duration. The detection rate or the non-detection of these transients by current and upcoming surveys can thus be used to place meaningful constraints on the extrapolation of the MBH mass function to moderate masses.
\end{abstract}

\section{Introduction} 

Stars that pass too close to a massive black hole (MBH) can be disrupted if the MBH's tidal gravitational field overwhelms the star's self-gravity. The approximate pericenter distance from the MBH for this to occur is the tidal radius, $\rt = (\Mbh / M_*)^{1/3} R_*$ \citep{Rees1988}. For the star to be disrupted, rather than swallowed whole, the tidal radius must be larger than the horizon radius of the MBH. For non-spinning MBHs, this implies $\rt \gtrsim \rs = 2 G \Mbh / c^2$. 
This paper examines transients generated by the tidal disruption of white dwarf (WD) stars, whose compactness means that the condition that  $\rt > \rs$ is satisfied only when  $\Mbh \lesssim 10^5 M_\odot$ \citep[e.g.][]{Kobayashi:2004kq,Rosswog:2009gg,MacLeod:2014gl,East:2014kl}.  Transient emission generated by tidal encounters between WDs and MBHs would point strongly to the existence of MBHs in this low-mass range. 

Identifying the distinguishing features of WD tidal disruptions as they might present themselves in surveys for astrophysical transients is of high value. These transients could help us constrain the highly-uncertain nature of the black hole (BH) mass function for BH masses between stellar mass and supermassive, just as main sequence star disruptions can do for supermassive BHs \citep{Magorrian:1999fd,Stone:2014vy}.
This paper focuses on the generation and detection of transients, and in so doing, it expands on a growing body of literature exploring the tidal disruption of WDs by MBHs \citep[e.g.][]{Luminet:1989wl,Kobayashi:2004kq,Rosswog:2008gc,Rosswog:2008hv,Rosswog:2009gg,Zalamea:2010eu,Clausen:2010vk,Krolik:2011ew,Haas:2012ci,Shcherbakov:2013hf,Jonker:2013es,MacLeod:2014gl,Cheng:2014jk,East:2014kl,Sell:2015wh}.  Two primary avenues for generation of bright transients have emerged from this work. First, dynamical thermonuclear burning can be ignited by the strong compression of the WD during its pericenter passage, producing radioactive iron-group elements. Secondly, in analogy to other tidal disruption events and their associated flares, the disruption of the WD feeds tidal debris back toward the MBH fueling an accretion flare.

\begin{figure}
\includegraphics[width=0.49\textwidth]{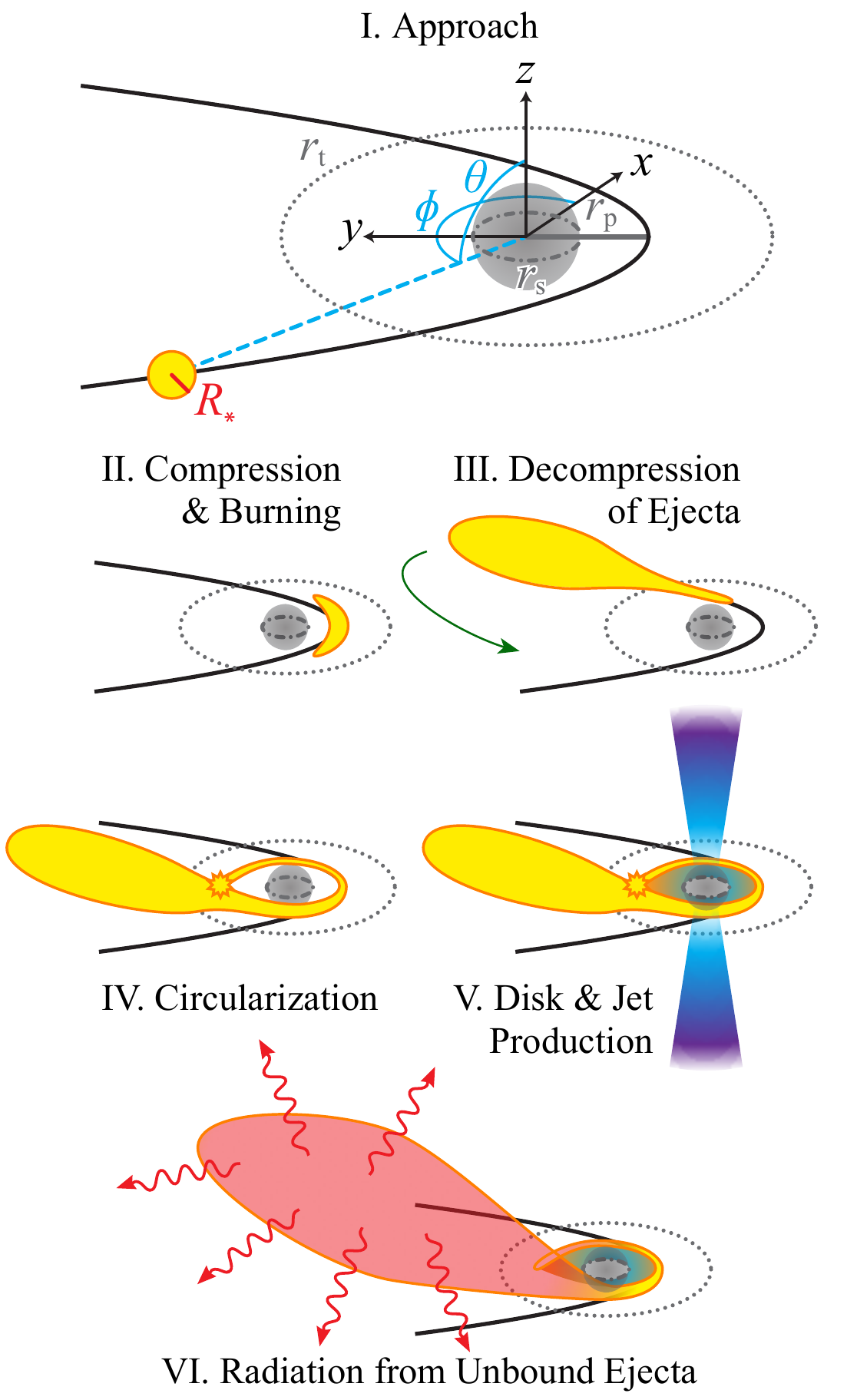}
\caption{
The sequence of events during a deep-passing WD tidal disruption events. A fraction of WD tidal disruption events pass sufficiently deeply within the tidal radius to ignite thermonuclear burning within the crushed WD material \citep{Rosswog:2009gg}. 
As bound debris falls back to the MBH and forms an accretion disk and jet, the unbound remnant material expands until radiation emerges due to decay of radioactive iron group elements in synthesized in the WD core. 
\label{Fig:diagram}}
\end{figure}

Thermonuclear transients from tidal compression of WDs can be generated in encounters where the WD passes well within the tidal radius at pericenter. The sequence of events in such an encounter is outlined in Figure \ref{Fig:diagram}. In these deep encounters, WD material is severely compressed in the direction perpendicular to the orbital plane (phase II in Figure \ref{Fig:diagram}), while at the same time being stretched in the plane of the orbit \citep{Carter:1982fn,Luminet:1985wz,Brassart:2008be,Stone:2012ul}. As portions of the star pass through pericenter, they reach their maximum compression and start to rebound \citep[phase III in Figure \ref{Fig:diagram} and, e.g. the hydrodynamic simulations of ][]{Rosswog:2008gc,Rosswog:2009gg,Guillochon:2009di}. 
 The thermodynamic conditions in the crushed WD material can be such that the local nuclear burning timescale is substantially shorter than the local dynamical timescale \citep{Brassart:2008be,Rosswog:2009gg}.
This implies that runaway nuclear burning can occur, injecting heat and synthesizing radioactive iron-group elements in the tidal debris.  The radiation takes some time to emerge from the debris, and the resultant transients are supernova (SN) analogs, with emergent radiation peaking at optical wavelengths, phase VI in Figure \ref{Fig:diagram} \citep{Rosswog:2008gc,Rosswog:2009gg}.  

In the meantime, bound debris falls back to the MBH. In order for accretion-fed transients to be generated, the tidal debris stream must fall back to the MBH and self-intersect to produce and accretion disk, shown as phase IV in Figure \ref{Fig:diagram}  \citep{Kochanek:1994bn,RamirezRuiz:2009gw,Dai:2013un,Hayasaki:2013kd,Bonnerot:2015vm,Hayasaki:2015ur,Shiokawa:2015wx,Guillochon:2015un,Piran:2015uk,2015arXiv150704333D}. If compact disks (with their proportionately short viscous draining times) do not form efficiently in all cases, then some events will produce luminous accretion flares while others do not \citep{Guillochon:2015un}.  The inferred accretion rates onto the MBH drastically exceed the MBH's Eddington accretion rate  $\dot M_{\rm Edd} = 4 \times 10^{-3} (\Mbh/10^5M_\odot)\ M_\odot \text{ yr}^{-1}$, where we have used $L_{\rm Edd} = 4\pi G \Mbh c / \kappa_{\rm es}$ and a 10\% radiative efficiency, $L=0.1 \dot M c^2$ ($\kappa_{\rm es}$ is the electron scattering opacity, 0.2 cm$^2$ g$^{-1}$ for hydrogen-poor material). This highly super-Eddington accretion flow, shown as phase V in Figure \ref{Fig:diagram}, may present an environment conducive to launching jets, implying that the observed signatures of accretion flare would be strongly viewing-angle dependent \citep{Strubbe:2009ek,Bloom:2011er,DeColle:2012bq,Metzger:2012hr,Tchekhovskoy:2013gw,Kelley:2014ff,MacLeod:2014gl}. Along a jet axis an observer would see non-thermal beamed emission, with isotropic equivalent luminosity similar to proportional to the accretion rate. Away from the jet axis, thermal emission would be limited to roughly the Eddington luminosity.  In both cases, the emission is likely to emerge at high energies with X-ray and gamma-ray transients \citep[e.g.][]{Komossa:2015uj} setting a precedent for their detection \citep{MacLeod:2014gl}.

This paper draws on simulations published by \citet{Rosswog:2009gg} to present detailed calculations of the emergent light curve and spectra from the radioactively powered thermonuclear transients. 
We present the hydrodynamic simulation method used and initial model in Section \ref{sec:model}, and the emergent optical lightcurves and spectra in Section \ref{sec:transients}. We discuss the process of WD debris fallback, accretion disk formation, and the associated accretion-fueled signatures of WD disruption in Section \ref{sec:accretion}. We then use our multiwavelength picture of transients emerging from close encounters between WDs and MBHs to discuss the prospects for the detection of thermonuclear transients at optical wavelengths and accretion signatures at X-ray and radio wavelengths by current and next generation surveys in Section \ref{sec:detect}. The primary uncertainty in the prevalence of these transients is the highly uncertain nature of the MBH mass function at low MBH masses.  In section \ref{sec:discussion} we discuss the implications of our findings with the ultimate goal of using WD tidal disruption transients as means to constrain the MBH mass function in the range of $10^3 - 10^5 M_\odot$. In Section \ref{sec:conclusion}, we summarize our findings and conclude.

\section{Hydrodynamic Simulation}\label{sec:model}

\citet{Rosswog:2009gg} performed detailed hydrodynamics-plus-nuclear-network
   calculations of the tidal compression of WDs with a particular focus on 
   those systems that lead to runaway nuclear burning.
 Our examination of the emergent lightcurves and spectra that define these transients is based on this work.   
The simulations of \citet{Rosswog:2009gg}  were carried out with  
   a smoothed particle hydrodynamics (SPH) code \citep{Rosswog:2008hv}.
   The initial stars were constructed as spheres of up 4 million SPH
   particles, placed on a stretched close-packed lattice so that equal-mass
   particles reproduce the density structures of the WDs. For both the initial
   WD models and the subsequent hydrodynamic evolution the {\tt HELMHOLTZ} equation
   of state \citep{Timmes:2000bq}. It makes no approximation in the treatment
   of electron-positron pairs and allows to freely specify the nuclear 
   composition that it can be conveniently coupled to nuclear reaction
   networks. The calculations used a minimal, quasi-equilibrium reduced reaction network
   \citep{Hix:1998fd} based on 7 species (He, C, O, Mg, Ne, Si-group, 
   Fe-group). Numerical experiments performed in the context of WD-WD 
   collisions \citep{Rosswog:2009wdc} showed that this small reaction network reproduces the energy 
   generation of larger networks to usually much better than 5\% accuracy.

The initial WD of 0.6 $M_\odot$ is cold ($T=5 \times 10^4$ K) and consists
   of  50\% carbon and  50\% oxygen everywhere \citep[Model 9 of][]{Rosswog:2009gg}.  This model is chosen since $0.6M_\odot$ WDs are among the most common single WDs \citep{Kepler:2007jz}.
 The model WD is subjected to a $\beta \equiv \rt/\rp = 5 $ encounter with a 500$M_\odot$ MBH.  
  \citet{Rosswog:2009gg} show that sufficiently deep passages lead to extreme distortions of the WD as it passes by the MBH.  In deep passages, the pericenter distance and the size of the WD may be comparable. As a result, the entire WD is not compressed instantaneously, instead a compression wave travels along the major axis of the distorted star  \citep[see, for example, Figure 6 of][]{Rosswog:2009gg}. 
For guidance, analytical scalings for polytropic stars suggest that at this point of maximum vertical compression, $z_{\rm min}/R_{\rm WD} \sim \beta^{-2/(\gamma-1)}$, where $\gamma$ is the polytropic index \citep{Stone:2012ul}.  Thus, inserting $\gamma=5/3$ gives the scaling  $z_{\rm min}/R_{\rm WD} \sim \beta^{-3}$ \citep{1986ApJS...61..219L,Brassart:2008be}. 
\citet{Rosswog:2009gg}'s three dimensional simulations show that the realities of a more complex equation of state and three dimensional geometry imply some departure from these scalings, but they nonetheless provide a useful framework in interpreting the compression suffered by close-passing WDs.

This compression results in nuclear burning along the WD midplane. 
Following the encounter, the inner core of $0.13 M_\odot$ 
has been burned completely to iron group elements, and is surrounded
by a shell of intermediate mass elements (IME) and an outer surface of
unburned C/O.  The compositional stratification resembles that of
standard SN~Ia models. The explosion has released an energy of
$10^{50.62}$~erg in excess of the gravitational binding energy of the initial WD. 
The synthesized  $0.13 M_\odot$  of iron-group
elements in the WD core are presumably in large part composed of radioactive
\Nifs.   Lacking detailed nucleosynthetic abundance calculations, however,
we must interpolate the abundances in order to compute
spectra and light curves.  In doing so, we are guided by
nucleosynthesis yields in calculations of SNe~Ia explosions by
\citet{1993A&A...270..223K}.  We assume that iron group material
consists of 80\% radioactive \Nifs, with the remaining 20\% being
stable isotopes (primarily $^{54}$Fe and $^{58}$Ni).  In regions of
IME we applied the compositions of the `O-burned' column of Table 1 of \citet{1999ApJS..121..233H}. Unburned carbon and oxygen material was assumed to be of solar metallicity.

\begin{figure}
\includegraphics[width=0.49\textwidth]{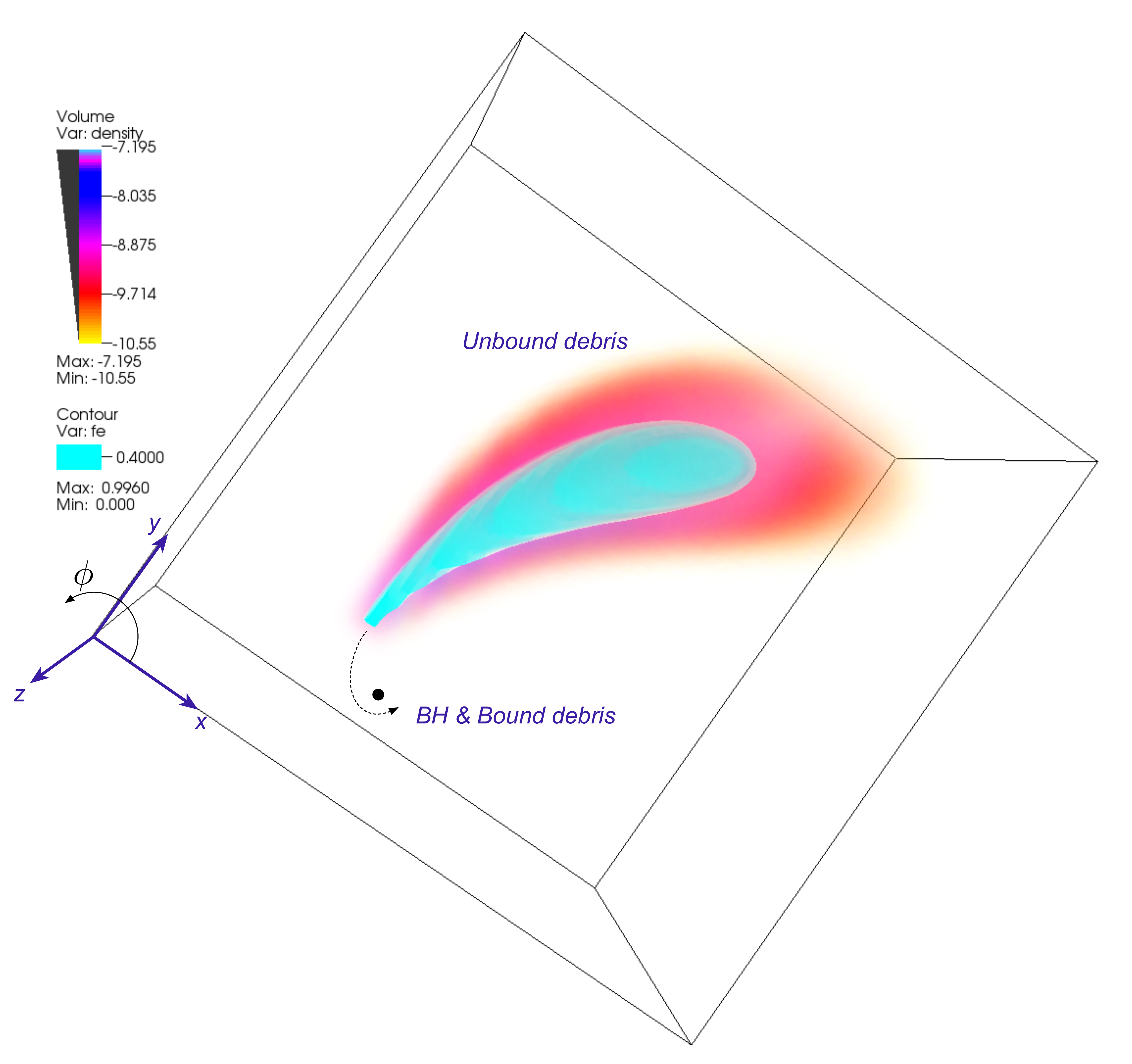}
\caption{
Visualization of the unbound remanent of the WD 
tidal disruption event. Red shading illustrates the density
structure of the $0.4 M_\odot$ of ejecta, while the cyan contour shows the iron group element
distribution. Coordinate directions are marked on one corner of the rendered volume (not at the coordinate origin).   
Angles in the plane of the orbit are defined as follows: an observer at $\phi = 0$ is aligned along the $\hat x$-direction, an observer along the $\hat y$-direction is at $\phi=\pi/2$. The initial pericenter approach is in the $-\hat y$-direction relative to the MBH, this is at $\phi=3\pi/2$.  In the orbital plane, viewing angles along the $y$-axis, $\phi= \sim \pi/2$ and $3\pi/2$, are surface area minimizing, while angles of $\sim 0$ and $\pi$ (along the $x$-axis) maximize the projected surface area. These differences are traced out in the observed brightness in Figure \ref{Fig:LC}.  
\label{Fig:ejecta}}
\end{figure}

Following the encounter, the subsequent hydrodynamical evolution of the unbound
material is computed for an additional 13.9 minutes as energy deposited by nuclear burning and gravitational interaction with the MBH shape the debris morphology.  
We show the structure of the unbound debris of the disrupted remnant in
Figure~\ref{Fig:ejecta}.  
Although the nuclear energy injection is significant, it does not dominate 
the energetics of the WD's passage by the MBH. 
The orbital kinetic energy at pericenter,
\beq
E_{\rm k, p} = {1 \over 2 } M_{\rm WD} v_{\rm p}^2  = {\beta \over 2} {G M_{\rm WD}^2 \over R_{\rm WD} } \left( \frac{\Mbh}{M_{\rm WD}} \right)^{2/3},
\eeq  
 is so large that even the $\sim 10^{51}$~ergs of explosion energy (similar to the WD binding energy) represents a perturbation
because this energy is larger than the WD's binding energy by a factor $\sim \left( \Mbh / M_{\rm WD} \right)^{2/3}$. 
 Because the orbital energy is so important, the morphology of the tidal debris 
largely retains its elongated shape seen in Figure \ref{Fig:ejecta}, and the explosion is far from spherical.
The energy injection does, however, shape the binding energy distribution of material to the MBH.
The final mass of unbound material is $0.4~\msun$, with the remaining
0.2~\msun\ of material expected to be accreted onto the MBH \citep[see Figure 4 of][]{Rosswog:2008gc}.  

The unbound remnant material moves along its hyperbolic 
orbital path with a bulk velocity of $9140$~\kms. 
The velocity of the least-bound debris following the  tidal encounter depends on the MBH mass, but only weakly,
$v_{\rm max} / v_{\rm esc} \approx  (\Mbh / M_{\rm WD})^{1/6}$, where $v_{\rm esc}$ is the WD escape velocity, $\sqrt{2 G M_{\rm WD}/R_{\rm WD} } $.
The nuclear and tidal energy injection lead to a 
maximum expansion velocity relative to the center of mass of about
12,000~\kms .  
At the end of the computation time (13.9 minutes after pericenter passage) the
mean gravitational (due to interaction with the MBH) and internal energy densities are small ($\la 1\%$)
relative to the kinetic energy density, and the velocity structure is
homologous ($v \propto r$) to within a few percent, indicating that
the majority of the remnant had reached the phase of approximate free-expansion.\footnote{In reality, despite the fact that the large majority of the material has obtained free expansion, 
an increasingly small portion of the marginally unbound material (binding energy near zero) continues to gravitationally interact with the MBH and we ignore this interaction in our analysis. }

\section{Optical Thermonuclear Transients}\label{sec:transients}

We calculate the post-disruption evolution of the WD remnant using a
3-dimensional time-dependent radiation transport code, {\tt SEDONA}, which
synthesizes light curves and spectral time-series as observed from
different viewing angles \citep{Kasen:2006et}. {\tt SEDONA} uses a Monte Carlo
approach to follow the diffusion of radiation, and includes a
self-consistent solution of the gas temperature evolution.  Non-grey opacities were used which
accounted for the ejecta ionization/excitation state and the effects
of $\approx 10$ million bound-bound transitions Doppler broadened by the
velocity gradients. We calculate the orientation effects by collecting
escaping photon packets in angular bins, using 30 bins equally spaced
$\cos\theta$ and 30 in $\phi$, where $\theta,\phi$ are the standard
angles of spherical geometry, defined relative to the plane of orbital motion (Figure~\ref{Fig:diagram}).
This provides the calculation of observables from 900 different viewing angles, each with equal
probability of being observed.

The diagnostic value of the
synthetic spectra and light curves in validating the ignition mechanism is striking. Light curve
observations constrain the energy of the explosion, the total ejected mass, and the amount and
distribution of $^{56}$Ni synthesized in the explosion. Spectroscopic observations at optical/UV/near-IR
wavelengths constrain the velocity structure, thermal state and chemical stratification of the ejected
matter. Gamma-ray observations constrain the degree of mixing of radioactive isotopes. Given the
 complexity of the underlying phenomenon, no single measure can be used to determine the
viability of an explosion model; instead, we will consider together a broad set of model
observables.

\subsection{Light Curves}
\label{Sec:LC}

The optical light curve of the remnant is powered by the radioactive decay
chain $\Nifs\rightarrow \Cofs \rightarrow \Fefs$ which releases $\approx
1$~MeV gamma-rays.  We model the emission, transport, and absorption
of gamma-rays which determine the rate of radioactive energy
deposition.  For the first 20 days after disruption, the mean free
path to Compton scattering is much shorter than the remnant size and,
despite the highly asymmetrical geometry, more than 90\% of gamma-ray
energy is absorbed in the ejecta.  This energy is assumed to be
locally and instantaneously reprocessed into optical/UV photons.

\begin{figure}
\includegraphics[width=3.3in]{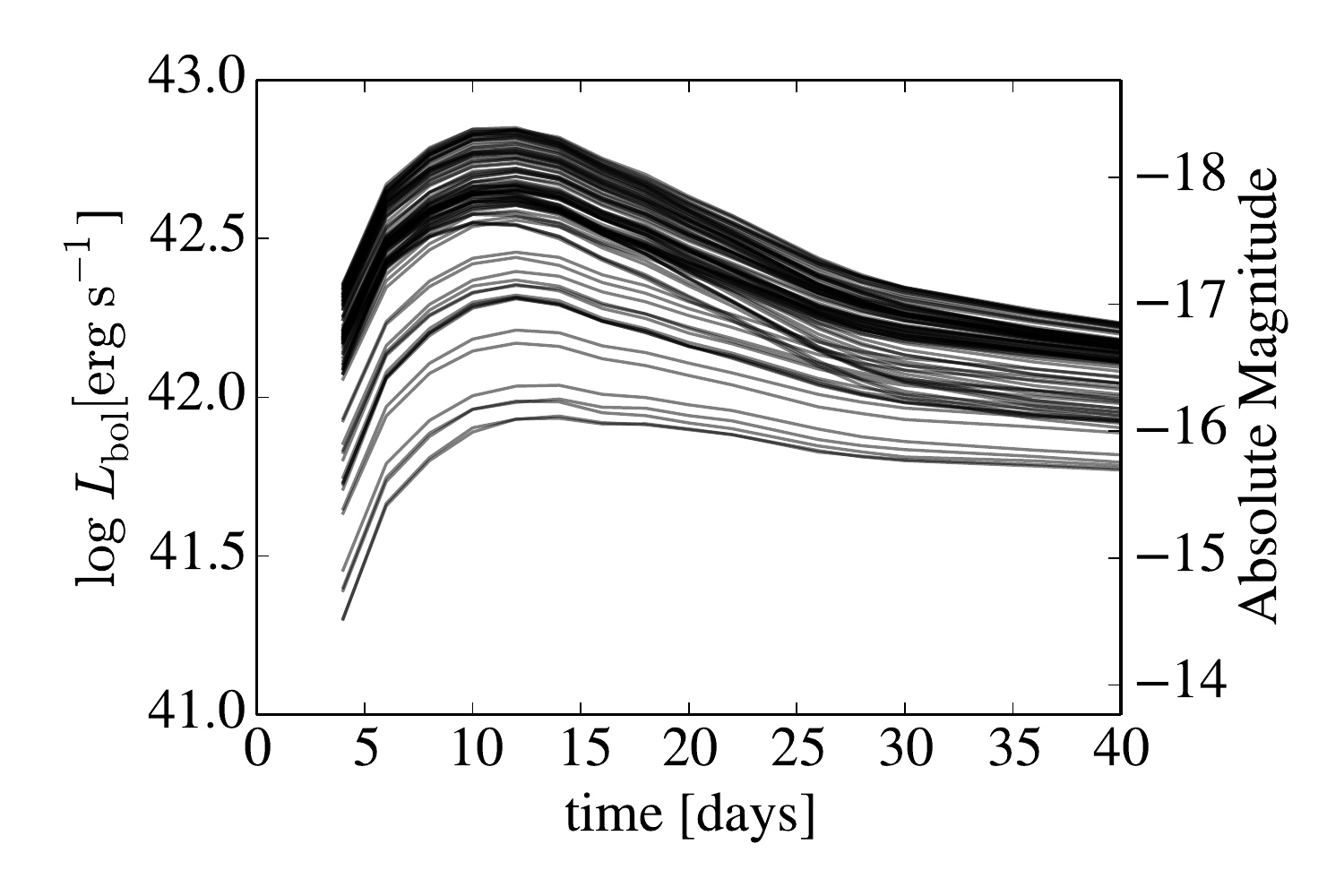}
\includegraphics[width=3.3in]{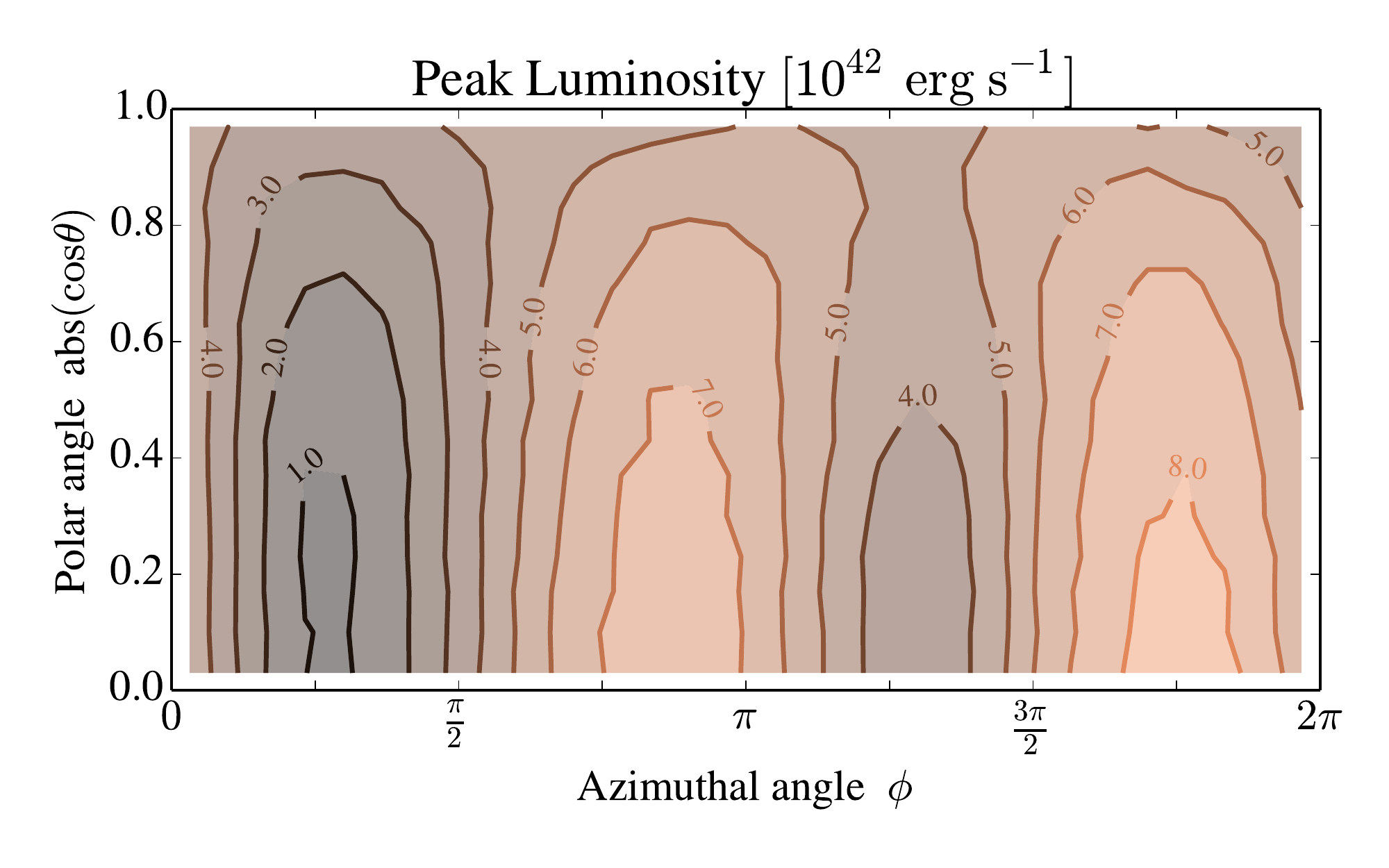}
\caption{The top panel shows the bolometric light curve of the model as observed
 from different viewing angles. Due to the asymmetric geometry of the ejecta (as seen in Figure \ref{Fig:ejecta}), the lightcurve peak and shape vary substantially with different viewing orientation. The lower panel shows the peak brightness as a function of viewing angle. The coordinates are defined such that the orbital plane is defined by  $\cos \theta = 0$. Since this is the plane in which the ejecta distribution and velocity changes with viewing angle, the peak luminosity changes by a factor $\sim 10$ with respect to changing $\phi$,  as mapped by the ejecta distribution shown in Figure \ref{Fig:ejecta}. 
  \label{Fig:LC}}
\end{figure}

In the upper panel of Figure~\ref{Fig:LC} we show the bolometric light curve
 as seen from various viewing angles.  The bolometric luminosity
reaches a peak about 10--12~days after the disruption. The rise time,
which depends little on the viewing angle, reflects the time-scale for
optical photons to diffuse from radioactively heated regions to the
remnant surface.  The diffusion time is shorter than for a Type~Ia
SNe (rise time of $\approx$18 days) because of the lower total ejected
mass and also the geometric distortion, which creates a higher surface area to volume ratio than for spherical ejecta.

The asymmetry of the remnant leads to strongly anisotropic emission.
The lower panel of 
Figure \ref{Fig:LC} maps the peak brightness (in units of $10^{42}$~erg~s$^{-1}$) 
onto the $\cos \theta$, $\phi$ viewing angle plane.
The peak luminosity varies by a factor of about $10$ depending on
orientation, with the model appearing brightest when its projected
surface area is maximized.  From such surface--maximizing orientations,
anisotropy enhances the brightness by a factor of several relative to
a comparable spherical model. The relatively short diffusion time also
promotes a higher luminosity.  The model therefore predicts maximum
peak luminosities of $\sim 8 \times 10^{42}$~erg~s$^{-1}$, similar to some
dim SNe~Ia making nearly 3 times as much \Nifs.  For the less common
surface--minimizing orientations, the model luminosity is significantly
reduced ($\sim 8 \times 10^{41}$~erg~s$^{-1}$).  These orientations may be compared to the coordinate system defined in Figure \ref{Fig:ejecta}.

\begin{figure}[tbp]
\begin{center}
\includegraphics[width=0.47\textwidth]{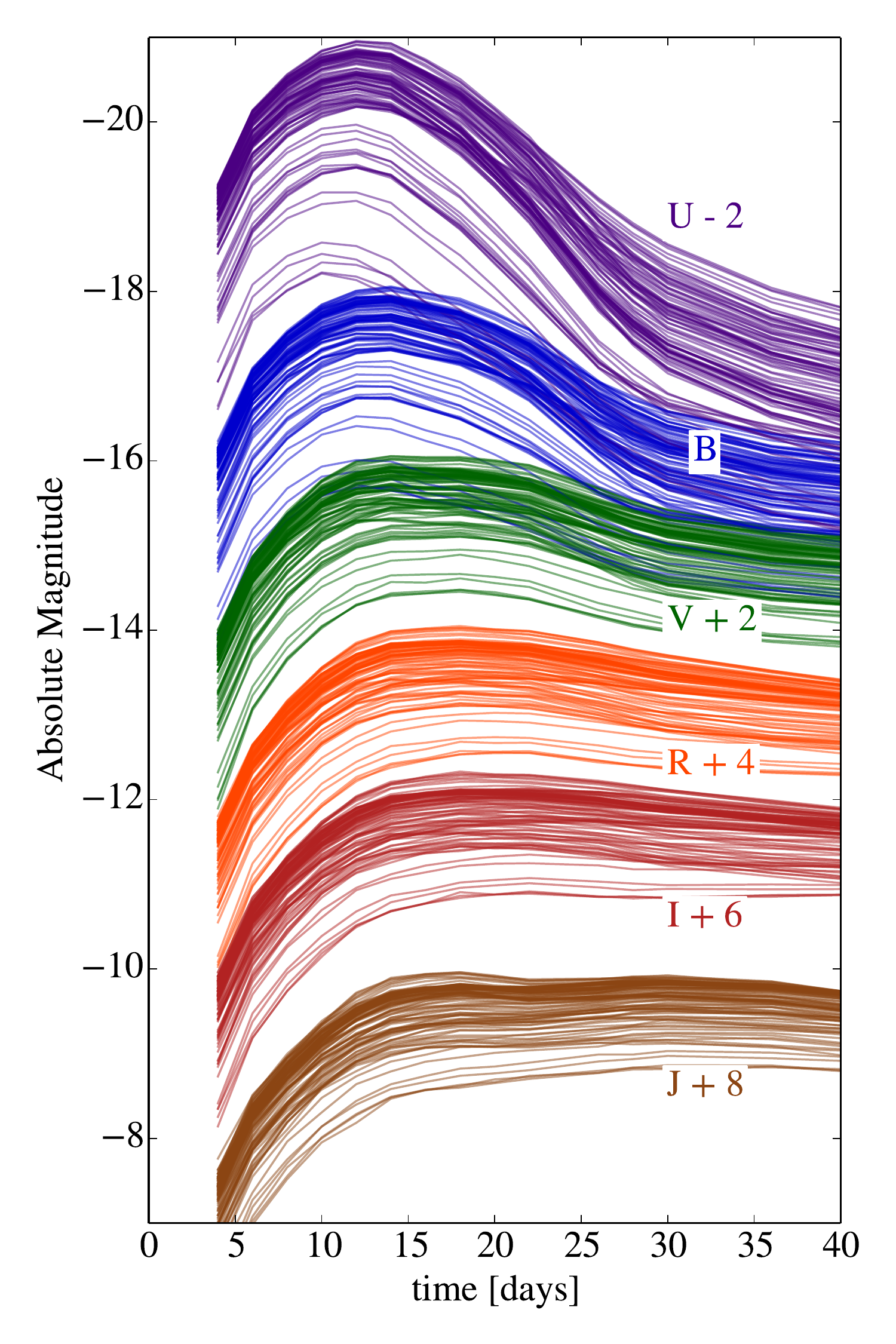}
\caption{Multiband lightcurves of the model thermonuclear transients realized from different viewing angles. 10 samples of $\cos \theta$ and 10 of $\phi$ are plotted, or a total of 100 different viewing angles. The U, B bands peak earliest, and show the strongest viewing angle dependence in time of peak and peak magnitude. Thus these blue colors best probe the early evolution, when the remnant is most asymmetric.  At later times, the dispersion between viewing angles narrows somewhat as the ejecta become increasingly spherical.  The I, J bands to do not show the strong decay then secondary maximum caused by recombination around day $\sim40$ of typical, luminous, type Ia SNe.  }
\label{fig:lc_color}
\end{center}
\end{figure}

In Figure \ref{fig:lc_color}, we decompose the bolometric lightcurve into Johnson UBVRIJ photometric bands by convolving the model spectra with filter transmission profiles. The absolute magnitudes are shown in Vega units. This figure shows several striking features of the thermonuclear transients. As one expects for typical thermonuclear SNe, the U and B bands are the most rapidly evolving. These peak earliest when the ejecta temperature is still quite high, and decline most rapidly from peak. The reddest bands, like I and J, do not show the strong second maximum caused by recombination that is observed in more luminous type Ia SNe \citep[e.g.][]{Kasen:2006et}.  These lightcurves also provide a quantitative measure of the dispersion in brightness in various filters with viewing angle. In general, the dispersion is broadest near peak and narrows at late times as the ejecta become increasingly spherical. Additionally, the viewing angle dependence is strongest in the U, B bands, due to differences in the degree of line blanketing of the blue part of the spectrum. We will explore this effect in more detail by examining the spectra in Section \ref{Sec:spec}.

\begin{figure}[tbp]
\begin{center}
\includegraphics[width=0.47\textwidth]{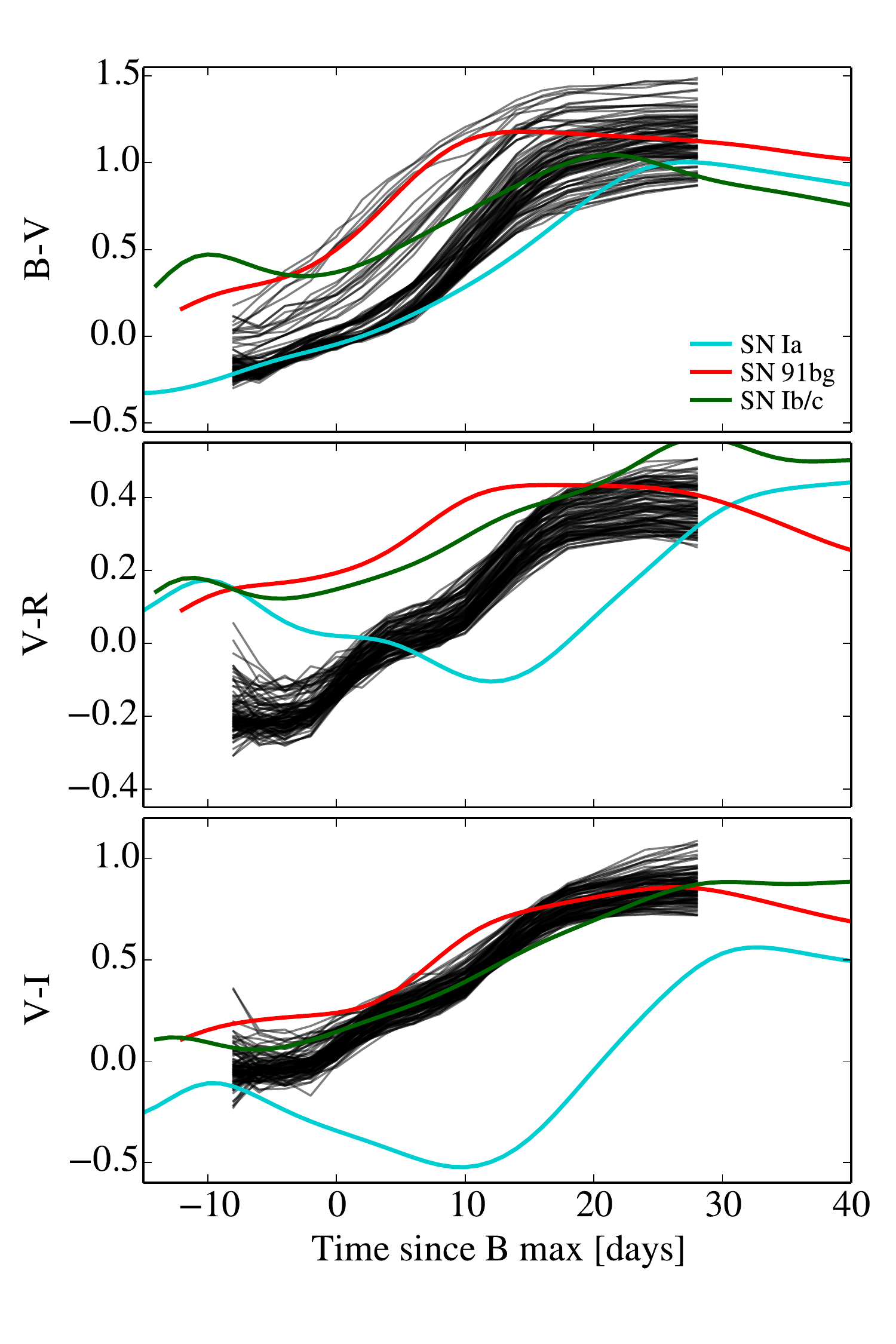}
\caption{Differential color evolution for our model thermonuclear transients at 100 different viewing angles evenly sampled in $\cos \theta$ and $\phi$. We compare to a selection of SN templates from P. Nugent described in the text.  A progressive reddening is observed in all colors as a function of time. The B-V curves show the broadest color spread with different viewing angles. Some angles show similar B-V evolution to typical Ia. V-R and V-I colors depart significantly from a standard type Ia, showing the lack of color inversion $\sim 10$ days after B maximum.  In general the color evolution is more consistent with less luminous events like the 1991 bg SNe and the type Ib/c. However, the V-R color  of the WD disruption transient is $\sim0.4$ magnitudes bluer than the more standard SNe, and evolves to redder colors more rapidly.}
\label{fig:lc_diff_color}
\end{center}
\end{figure}

Another useful photometric diagnostic of the multicolor lightcurves is the evolution in relative colors. We examine B-V, V-R and V-I lightcurves of our models (again sampled over 100 viewing angles) in Figure \ref{fig:lc_diff_color}. 
For comparison, we overplot differential colors from P. Nugent's lightcurve templates\footnote{ {\tt https://c3.lbl.gov/nugent/nugent\_templates.html } } for SNe of normal type Ia, 1991bg-like  \citep{2002PASP..114..803N}, and SN Ib/c \citep{2005ApJ...624..880L} categories. 
All colors show consistent reddening in time over the duration for which we have model data.  The B-V color shows the most variability with viewing angle, with a spread of $\sim1$ magnitude shortly after peak (e.g. 0--10 days in Figure \ref{fig:lc_diff_color}).  By contrast, the V-R and V-I colors show relatively narrow dispersion, despite the ejecta anisotropy. This can likely be traced to differences in the opacity source for these respective wavelengths. While the V, R, and I bands are largely electron-scattering dominated, the B-band is subject to strong Fe-group absorption features. 
The WD disruption transient's colors deviate from those of more common SNe in the V-R color for 1991bg events and type Ib/c and in both the V-R and V-I colors for normal type Ia.

\begin{figure}[tbp]
\begin{centering}
\includegraphics[width=0.49\textwidth]{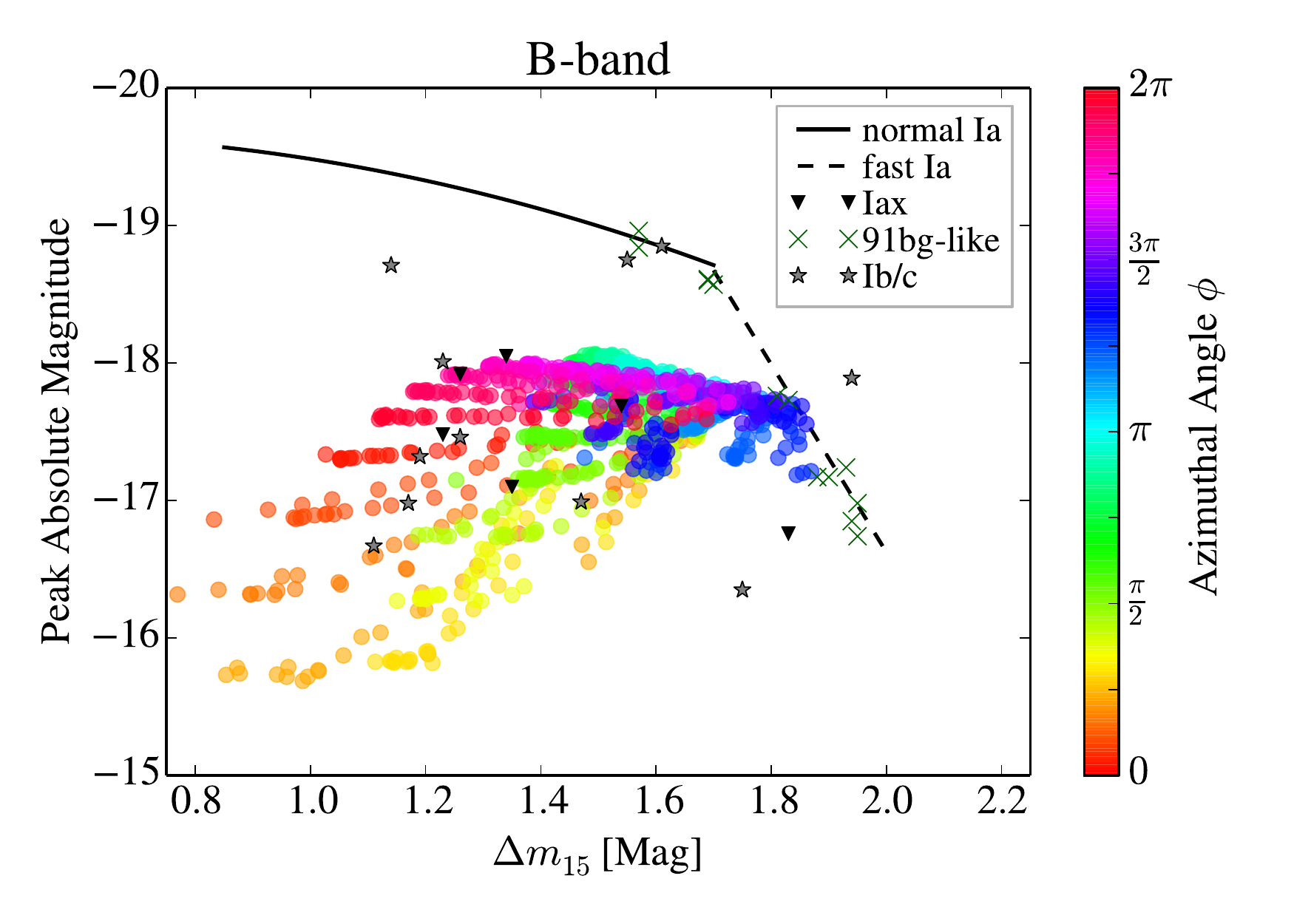}
\caption{Position of of model thermonuclear transients in width-luminosity phase space \citep{Phillips:1993he}. Model transients exhibit a wide range of lightcurve peak magnitudes and $\Delta m_{15}$ in the B-band. The most luminous with $M_B\approx -18$ exhibit a range of $1\lesssim \Delta m_{15} \lesssim 1.9$  Standard, luminous Ia's are plotted as a black line showing the \citet{Phillips:1999fm} relationship. Fast and underluminous Ia SNe are shown as a dashed line with the relationship of \citet{Taubenberger:2008df}. Data for a sample of representative 91bg-like, Iax, and Ib/c SNe are shown for comparison \citep[sample data from][Figure 5]{Drout:2013jn}. 
The tidal thermonuclear transients exhibit similar widths to standard Ia, despite their fainter peak magnitude. This puts them in similar width-luminosity phase space to SNe like Ib/c, and Iax. 
 }
\label{fig:phillips}
\end{centering}
\end{figure}

In Figure \ref{fig:phillips}, we examine the lightcurve shape and how it varies with viewing angle by plotting the B-band peak magnitude and $\Delta m_{15}$ \citep{Phillips:1993he}. We also plot the \citet{Phillips:1999fm} relation (with a solid line) for normal type Ia and the \citet{Taubenberger:2008df} relationship for 91bg like SNe (dashed line).  Overplotted on this space are a selection of type Iax, 91bg like, and Ib/c SNe \citep[with sample data provided by M. Drout from Figure 5 of][]{Drout:2013jn}.  Many of the tidal thermonuclear transients occupy similar phase space to SN type Iax and Ib/c -- exhibiting similar decline rates but lower luminosities as compared to normal Ia. They are photometrically distinct however, from normal Ia and 91 bg SNe. The fastest-declining viewing angles, with $\phi \sim 4$, extend to the tail of the quick-declining tail of the normal SNe distributions. 

\subsection{Spectra}
\label{Sec:spec}

In this section we explore the distinctive features of synthetic spectra of the WD tidal disruption thermonuclear transient. 
The model spectra of a tidally disrupted WD broadly resembles those of SNe,
with  P-Cygni feautres superimposed on a pseudo-blackbody continuum.

\begin{figure}
\centering
\includegraphics[width=0.49\textwidth]{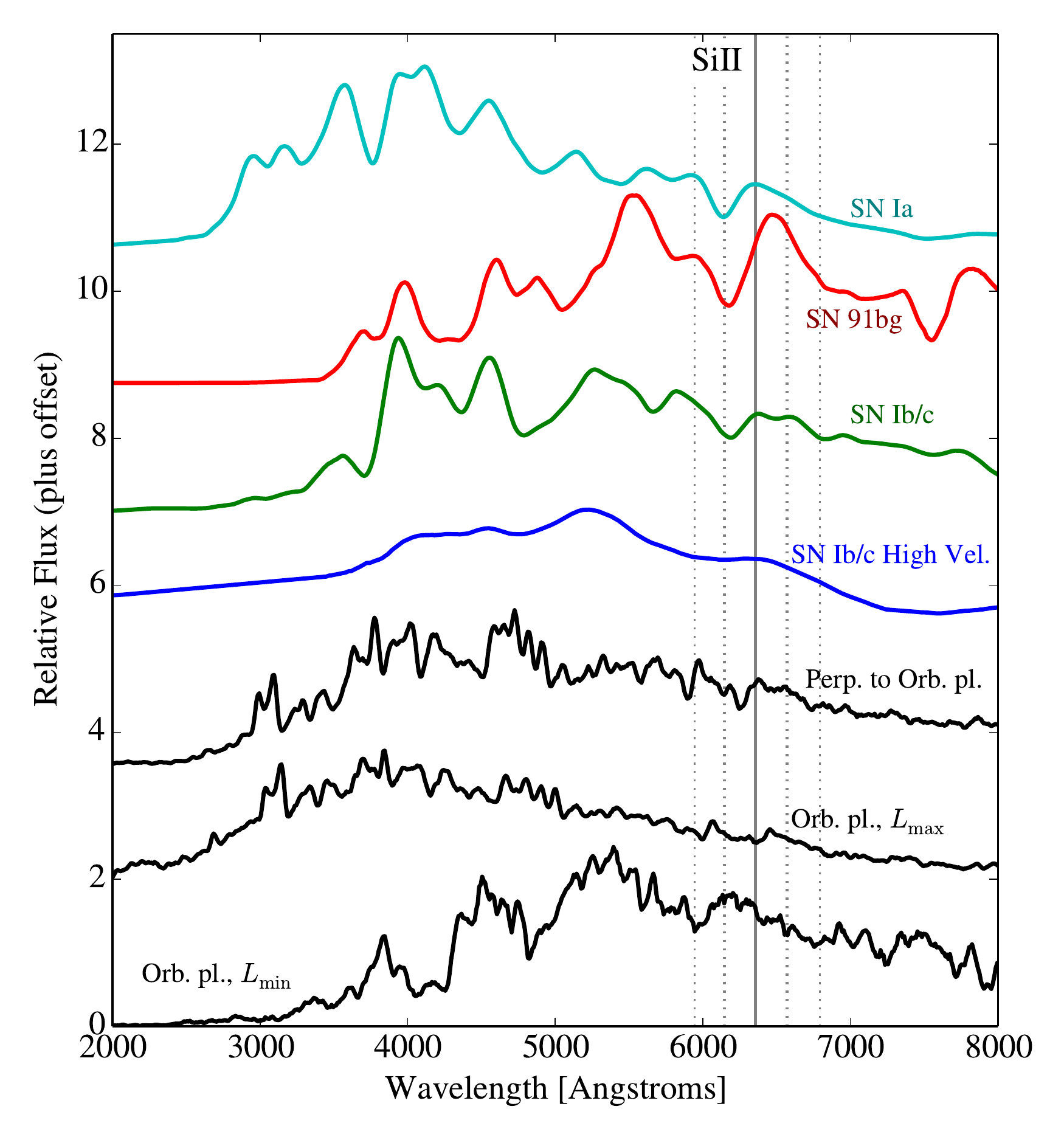}
\caption{
Spectra at $t=20$ days viewed in the orbital plane ($\cos \theta = 0$) (black, lower two spectra), the upper spectrum is near the viewing angle of maximum brightness $\phi \approx 5.5$, while the lower spectrum is near the minimum brightness $\phi \approx 1$.  A spectrum viewed perpendicular to the orbital plane is also plotted. The blue wavelengths of the minimum brightness spectrum are heavily blanketed by broad and strongly blue-shifted lines. By contrast, the upper spectrum, from the brighter viewing angle is bluer and exhibits very narrow line features near their rest wavelength. The narrow lines as seen from this orientation create a much lower effective opacity in this direction and higher photosphere effective temperature as the observer sees deeper into the ejecta.    Model spectra are smoothed with a 50 Angstrom rolling average. These spectra are compared to a spectral templates at $t=22$ days for SN type Ia, SN 91 bg, SN Ib/c and a high-velocity SN Ib/c \citep[from P. Nugent's online templates, original sources:][]{2002PASP..114..803N,2005ApJ...624..880L}. 
The minimum brightness spectrum in the orbital plane shares some similarities with SNe of type 1991 bg, which share relatively low nickel masses. However, it is quite distinct from the type Ib/c and high velocity type Ib/c events that are often associated with gamma ray bursts. 
 \label{Fig:minmaxspec}}
\end{figure}

Figure~\ref{Fig:minmaxspec} compares representative model spectra at
day 20 to those of several characteristic SNe classes. 
We plot the spectrum perpendicular to the orbital plane, and in the orbital plane along the directions of 
maximum and minimum brightness (as mapped in the lower panel of Figure \ref{Fig:LC}). 
The model spectra show common features of intermediate mass elements, 
in particular Si~II and Ca~II, and a broad absorption near 4500~\AA\ due
 to blended lines of Ti~II, which is characteristic of cooler, dimmer SNe \citep[e.g.][]{2002PASP..114..803N}.
Figure~\ref{Fig:minmaxspec} also shows that the spectrum of the model varies
significantly with orientation.  The absorption features are weaker
and narrower from ``face-on'' viewing angles (relative to the distorted ejecta structure), as the velocity gradients along the line of sight are minimized.  
These angles include perpendicular to the orbital plane, and in the orbital plane where the surface area is maximized (orientations of maximum brightness). 
Line absorption features are stronger from pole-on views.  Similarly, the
degree of line blanketing also varies with viewing angle, which
affects the color of the transient.  In general, the models have more
blanketing in the blue and UV and so look redder from pole-on views.

Immediate distinctions between the representative SNe categories and the WD tidal disruption model spectra are visible in Figure \ref{Fig:minmaxspec}. 
The model spectra share relatively narrow lines of SN Ia and SN 91 bg \citep[data from P. Nugent's online templates, which have original sources of][]{2002PASP..114..803N,2005ApJ...624..880L}. But the orientations perpendicular to the orbital plane and along the direction of maximum luminosity show much weaker line signatures than do our model spectra, which, at these orientations, only show weak line features. 
The orientation of minimum luminosity shares more spectral commonalities with the SN 91 bg spectra, especially the blended absorption features seen in the blue. 
The model spectra are, however, quite distinct from SN Ib/c spectra and the high velocity Ib/c spectra typically associated with gamma ray bursts (GRBs). Both of these Ib/c template spectra show significantly broader features than all but the most extreme (and minimum luminosity) WD disruption transient spectrum. 

\begin{figure*}[tbp]
\begin{centering}
\includegraphics[width=0.8\textwidth]{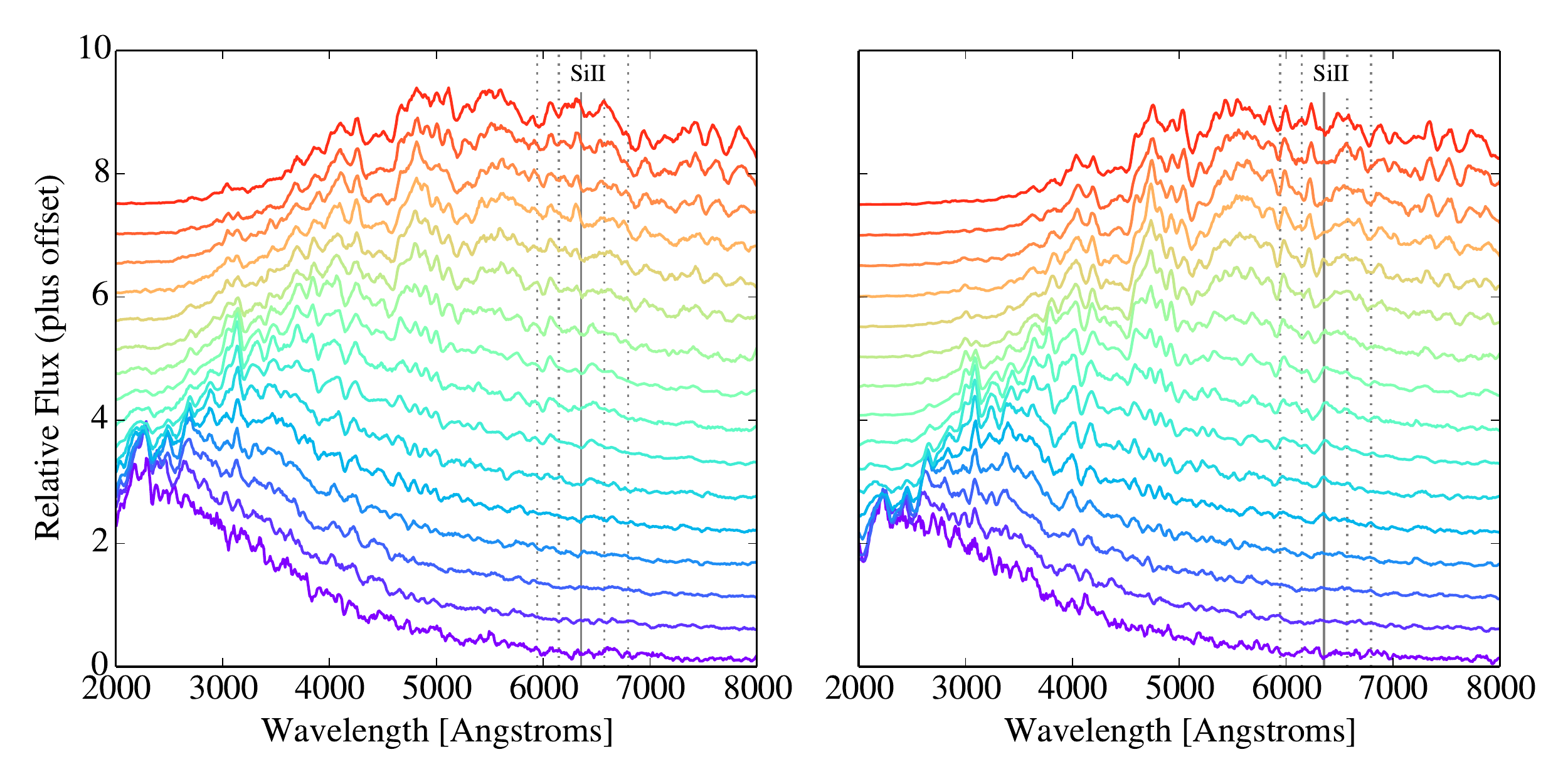}
\caption{Time series of spectra for two different viewing angles. On the left, $\cos \theta \approx 0$, $\phi \approx 5.5$.  On the right, $\cos \theta \approx 0.84$, $\phi \approx 2.8$. Thus, the left timeseries is viewed in the orbital plane near the orientation of maximum brightness, while the right is viewed close to face on. The timeseries consists of spectra at times of $t = $4, 6, 8, 10, 12, 14, 16, 18, 20, 22, 24, 26, 28, 30, 36, and 40 days (violet to red). The basic evolution from blue to redder continuum is accompanied by the development of strong and broad P-Cygni lines.  Solid lines show the rest wavelength of the SiII 6355 feature, while dashed and dotted lines show a $10^4$ km s$^{-1}$ and $2 \times 10^4$ km s$^{-1}$ velocity offsets, respectively. Model spectra show some Monte Carlo noise, which is smoothed with a 50~\AA\ rolling average. 
 \label{Fig:timeseriesspec}}
\end{centering}
\end{figure*}

Figure \ref{Fig:timeseriesspec} shows the spectral evolution of the remnant 
from two viewing angles. The viewing angle in the left panel is in the plane of WD orbit about the MBH, $\cos \theta = 0$. We choose the azimuthal angle of maximum brightness, $\phi \approx 5.5$.  The right-hand panel shows the spectral evolution as viewed nearly perpendicular to the orbital plane, where  $\cos \theta = 0.84$, or $\theta \approx 30^\circ$.  At this second orientation there should be less variation with azimuthal angle. 
Spectra are plotted at times of $t = $4, 6, 8, 10, 12, 14, 16, 18, 20, 22, 24, 26, 28, 30, 36, and 40 days after pericenter passage (as the line color goes from violet to red). This timeseries shows the progressive reddening of the spectral energy distribution as well as the gradual emergence of blended absorption lines in the spectrum.  Strong absorption features due to intermediate mass elements only appear around day $\sim 14$, several days after peak brightness, while the early spectra are relatively featureless.  This can be contrasted to typical type Ia, which show strong P-Cygni lines near peak.  Model spectra also show some Monte Carlo noise, which we have smoothed by adopting a 50~\AA\ rolling average of the model data's 10~\AA\ bins.

\begin{figure}[tbp]
\begin{centering}
\includegraphics[width=0.49\textwidth]{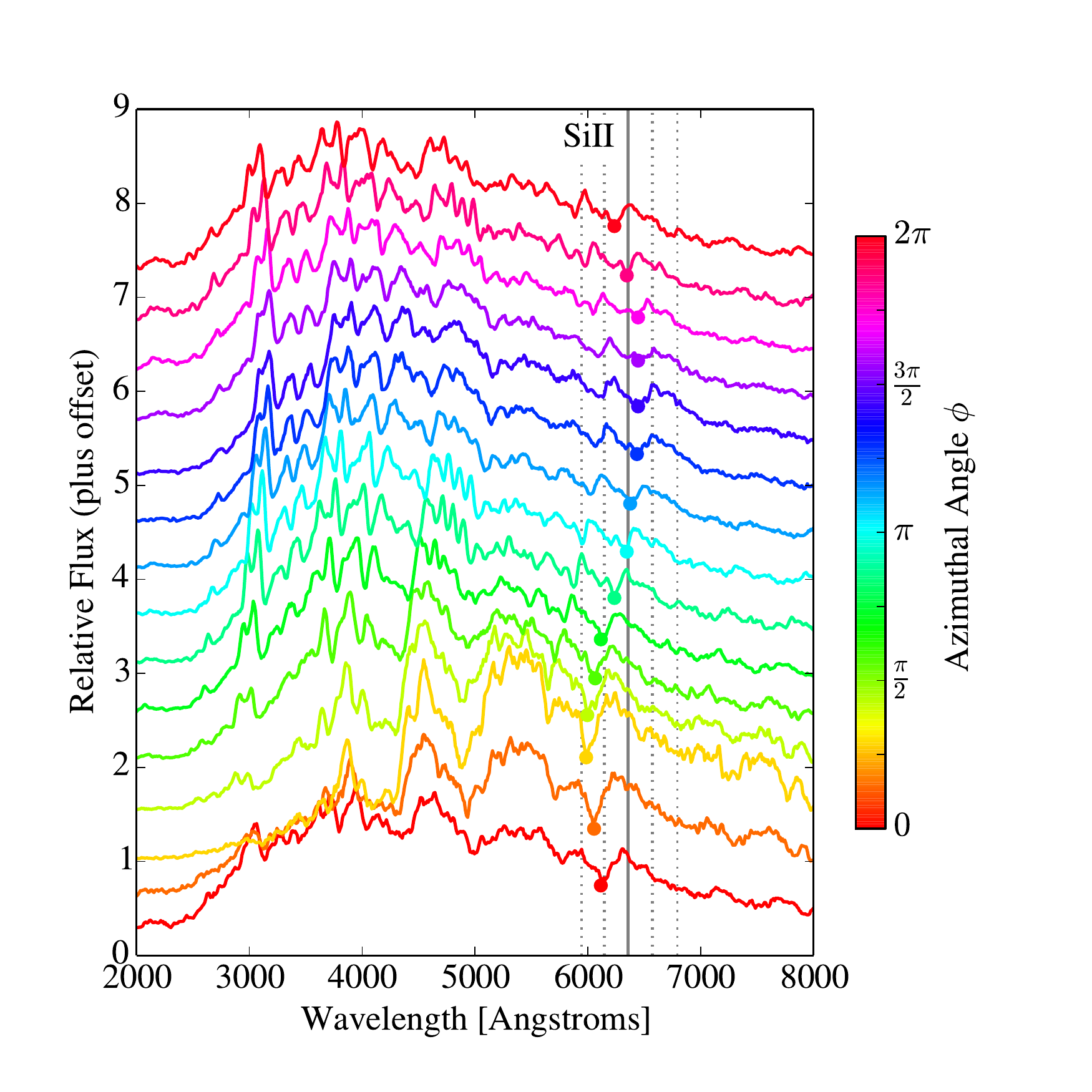}
\caption{
Spectra at $t=20$ days, as viewed from $\theta = 60^\circ$ ($\cos \theta = 0.5$), and $\phi$ ranging from 0 to $2 \pi$.  As viewing angle varies, the Doppler shift of the P-Cygni line features varies significantly with viewing angle. The ejecta is maximally blue-shifted near $\phi \approx 1$ and maximally red-shifted near $\phi \approx 4$, offset by a factor of $\pi$ in viewing angle. 
Spectra are smoothed with a 30~\AA\ rolling average. A point marks the minimum of the SiII 6355 feature as viewed from each azimuthal angle. This can be compared with the dashes in background which show  $10^4 $km s$^{-1}$ and $2 \times 10^4$ km s$^{-1}$ velocity offsets from the rest wavelength. 
 \label{Fig:anglespec}}
 \end{centering}
\end{figure}

When viewed from altitude orientations other than $\cos \theta = \pm 1$, the projection of the hyperbolic orbital motion of the unbound debris of the tidal disruption event plays a role in shaping the spectra with respect to viewing angle. 
Figure \ref{Fig:anglespec} shows the variance in spectra with viewing angle $\phi$, the azimuthal angle in the plane of the orbit with $\cos \theta = 0.5$ at t~=~20 days. One can immediately see that the orbital motion of the WD offers an important signature for tidal
disruption events.  The velocity dispersion due to expansion from the center of
mass produces the typical broad P-Cygni absorption profile, with
a line minimum blueshifted by $\sim 10,000~\kms$.  However, the bulk
orbital motion contributes an additional overall Doppler shift of
comparable magnitude.  From viewing angles in which the remnant is
receding from the observer, the two velocity components nearly cancel
and the absorption feature minima occur near the line rest wavelength.  
From viewing angles in which the remnant
is approaching the observer, the two velocity components add and the
blueshifts reach $\sim 20,000~\kms$.  

\begin{figure}[tbp]
\begin{centering}
\includegraphics[width=0.49\textwidth]{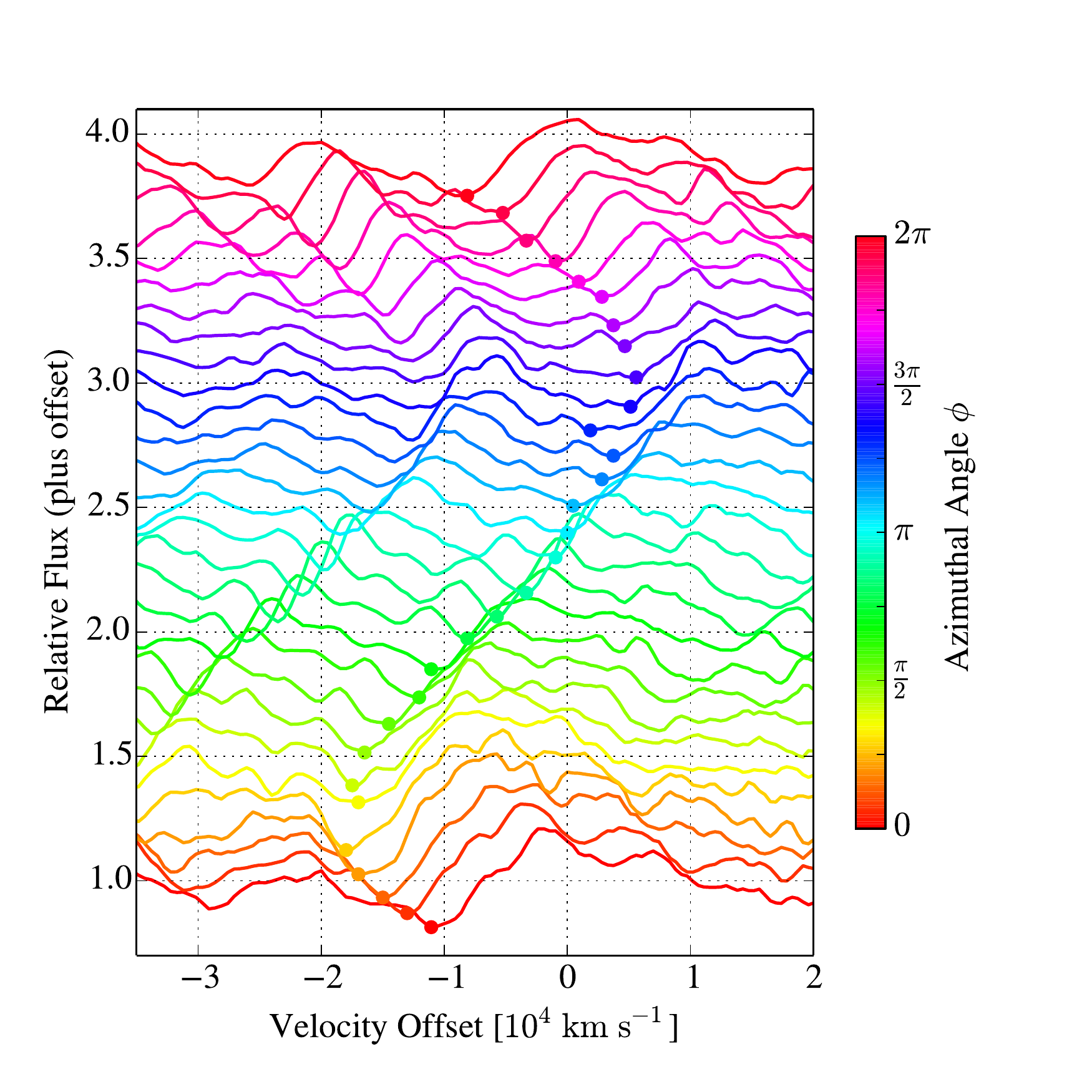}
\caption{
Normalized model spectra near the SiII 6355 feature from the same time and viewing angles ($t=20$ days,  $\theta = 60^\circ$) as Figure \ref{Fig:anglespec}. In this panel we plot all 30 viewing angles sampled by the radiative transfer calculation. The x-axis shows velocity offset from the rest wavelength of the SiII feature (6355 Angstroms).  The maximum absorption of the line feature shifts from +5000 km s$^{-1}$ (redshift) to -18,000 km s$^{-1}$ (blueshift) depending on azimuthal viewing angle. These spectra are each normalized by an approximate continuum constructed through a 1000~\AA\ blend of the model spectrum, then smoothed with a 40~\AA\ rolling average.. 
 \label{Fig:veloffset}}
 \end{centering}
\end{figure}

In Figure \ref{Fig:anglespec} we explore these Doppler shifts by marking the point of maximum absorption in the SiII 6355 feature with a point. One can see this makes a characteristic "S" curve with semiamplitude of $\sim 10^4$ km s$^{-1}$ as seen from this viewing angle and time. 
In Figure \ref{Fig:veloffset}, we zoom in on the region around the SiII 6355 feature, and plot the line profiles in velocity space for all 30 azimuthal viewing angles with $\cos \theta = 0.5$. In this panel we plot spectra that are normalized by the continuum constructed by a 1000~\AA\  blend of the spectrum.  In these line profiles, one sees an absorption feature with width $\sim 8000 - 10,000$ km s$^{-1}$.   As we scan through viewing angle, the line's P-Cygni profile shifts relative to its rest wavelength with the bulk orbital motion of the WD. Some variation in line strength and shape is also visible as we view the highly asymmetric ejecta structure from a variety of angles.

We can conclude that the discovery of a transient with
reasonably broad SN-Ia like absorptions, but with spectral systematically
off-set from the host-galaxy redshift by 5000-10,000~\kms\ would be
strong evidence of tidal disruption event.

\section{Multi-wavelength Accretion Counterparts}\label{sec:accretion}

In addition to the photometric and spectroscopic features described above, 
thermonuclear transients arising from tidal disruptions of WDs are unique among the diverse zoo of stellar explosions in that they are accompanied by the accretion of some of the tidal debris onto the MBH. 
In this section we describe predictions for the accompanying accretion signatures.

\subsection{Debris Fallback, Circularization, and Accretion}

Following a tidal disruption of a star by a MBH, a fraction of the material falls back to the vicinity of the  hole. This gravitationally bound material can power a luminous accretion flare \citep{Rees1988}.  In the case of tidally triggered thermonuclear burning of a WD, the injection of nuclear energy into the tidal debris does change the mass distribution and fraction of material bound to the MBH by tens of percent \citep{Rosswog:2009jc}. 
The traditional thinking has long been that material falling back to the vicinity of the MBH will promptly accrete subsequent to forming a small accretion disk with size twice to the orbital pericenter distance \citep{Rees1988}. This long-held assumption has recently been called into question by work that has examined the self-intersection of debris streams in general relativistic orbits about the MBH \citep{Dai:2013un,Hayasaki:2013kd,Bonnerot:2015vm,Hayasaki:2015ur,Shiokawa:2015wx,Guillochon:2015un,Piran:2015uk,2015arXiv150704333D}. Self-intersection, and thus accretion disk formation, are found to be much less efficient than originally believed, potentially slowing the rise and peak timescales of many tidal disruption flares \citep{Guillochon:2015un}, while also potentially modifying their characteristic flare temperatures \citep{2015arXiv150704333D}.

\begin{figure}[tbp]
\centering
\includegraphics[width=0.49\textwidth]{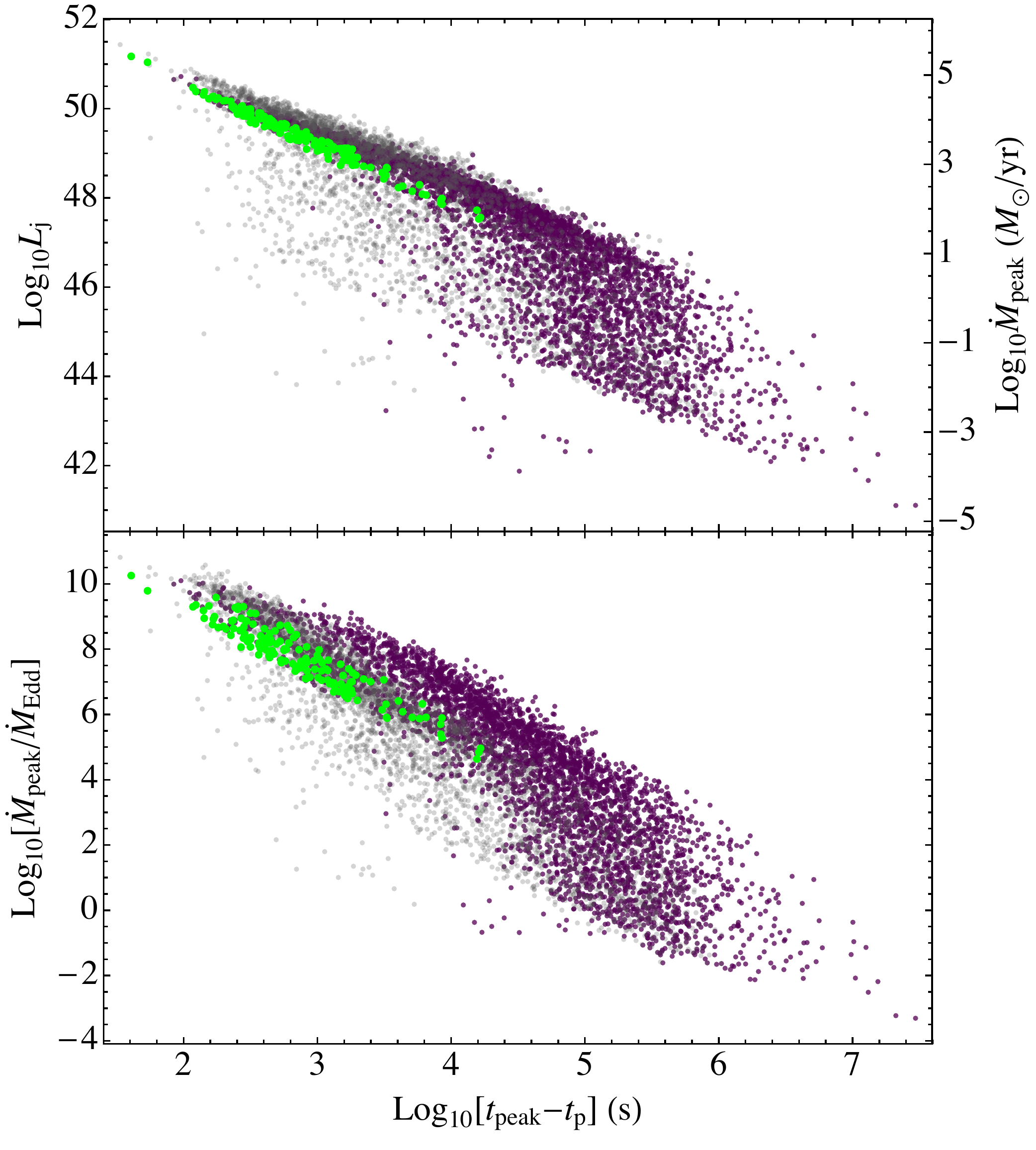}
\caption{Peak timescales and accretion rates for WD disruptions by MBHs with masses $10^2 M_\odot < \Mbh < 10^7 M_\odot$. WD masses are chosen from a distribution around $0.6 M_\odot$, and combinations in which the WD would be swallowed whole are rejected.  In the lower panel, accretion rates are plotted relative to the MBH's Eddington rate, and in solar masses per year  on the right-hand axis of the upper panel.  Illustrative jet luminosities are generated assuming an efficiency of 10\%, $L_j=0.1 \dot M c^2$ \citep[e.g.][]{DeColle:2012bq,Krolik:2012da} on the left-hand axis of the upper panel. Grey points show the intrinsic fallback rates \citep{Guillochon:2013jj}, while purple points show the timescales slowed or delayed by the debris stream's self-intersection time upon return to pericenter \citep{Guillochon:2015un}. Green points highlight events with $\beta > \beta_{\rm thermo}$ that would lead to thermonuclear ignition and the associated transient.  Those events that lead to thermonuclear transients are preferentially prompt and generate highly super-Eddington accretion rates with peak timescales of $10^2-10^4$ s after pericenter passage, offering promising prospects for precursor emission.}
\label{Fig:accretion}
\end{figure}

We explore the effects of debris circularization on flare peak accretion rates and timescales for WD tidal disruptions in Figure \ref{Fig:accretion}. We use the same approach as \citet{Guillochon:2015un} and perform Monte Carlo simulations of tidal disruptions and their circularization, but we use parameters appropriate for WD disruptions. In this model, the star is assumed to be initially on a zero-energy (i.e. parabolic) orbit, with its periapse distance being drawn assuming a full loss cone, $P(\beta) \propto \beta^{-2}$ \citep{Magorrian:1999fd,MacLeod:2014gl}. We sample WD masses from a distribution around $0.6 M_\odot$, $P(M_{\rm WD}) \propto \exp (-0.5 [(M_{\rm WD}/M_\odot-0.6)/0.2]^2 )$   \citep[e.g.][]{Kepler:2007jz}. Because the distribution of MBH masses is uncertain for $M_{\rm h} \lesssim 10^{6} M_{\odot}$ \citep{2008ApJ...688..159G}, $M_{\rm bh}$ is drawn from a flat distribution in $\log \Mbh$ between $10^2 M_\odot$ and $10^7 M_\odot$. Black hole spin $a_{\rm spin}$ is drawn from a flat distribution ranging from zero to one. We reject combinations of WD and MBH mass that would lead to the WD being swallowed whole by the MBH by redrawing $M_{\rm WD}$, $\beta$ and $a_{\rm spin}$; in cases which fail to produce a valid disruption after 100 redraw attempts, a new MBH mass is chosen (this allows us to include in our sampling the small fraction of events that are not swallowed by MBHs with $\Mbh \gtrsim 10^6 M_\odot$).

The peak accretion rate $\dot{M}_{\rm peak}$ and time of peak $t_{\rm peak}$ are determined for each disruption using the fitting formulae of \citet{Guillochon:2013jj}, assuming that WDs with mass $M_{\rm WD} < M_{\odot}$ are $\gamma = 5/3$ polytropes and $M_{\rm WD} > M_{\odot}$ are $\gamma = 4/3$ polytropes. In Figure \ref{Fig:accretion}, the underlying grey points show the distributions in these quantities realized by the prompt fallback rate to the MBH and the timescale of peak relative to pericenter passage.  Purple points show full and partial disruptions with $\beta < \beta_{\rm thermo}=3$. Green points show events that will also lead to the thermonuclear burning of the WD, where $\beta > \beta_{\rm thermo}$. 

Comparing the fallback and accretion rates suggests that tidal disruption flares resulting from WD encounters with MBHs are slowed in timescale and reduced in peak accretion rate by up to one order of magnitude due to circularization effects (As seen in the upper panel of Figure \ref{Fig:accretion}). The typical ``dark period,'' the time between disruption and when the stream first strikes itself, is found to be small, with a median value of just $1.6\times10^3$~s. This is roughly the amount of time the most-bound debris takes to complete one orbit, which is different than what is typical for main-sequence star disruptions by supermassive BHs where the stream can wrap a dozen times around the BH before striking itself \citep{Guillochon:2015un}. While WD disruptions are likely to be just as relativistic as their main sequence disruption cousins \citep{RamirezRuiz:2009gw}, the debris stream size is larger compared to the pericenter distance due to the smaller mass ratio of star to MBH, this means that the small deflection introduced by MBH spin is less capable of facilitating a long dark period.

With the factor of ten reduction in accretion rate due to inefficient circularization, WD disruption flares are somewhat less luminous than one would determine from the fallback rate. However, the lower panel of Figure \ref{Fig:accretion} demonstrates that nearly all WD tidal disruptions generate super-Eddington peak accretion rates, with many leading to accretion many orders of magnitude above the Eddington limit, even after accounting for the viscous slowdown of the flares. Events that lead to tidal thermonuclear transients nearly generate accretion flows onto the MBH that range from $10^5 - 10^{10}$ times the MBH's Eddington rate. 

\subsection{Accretion Disk and Thermal Emission}

Despite the peak mass fallback rate greatly exceeding the Eddington limit, we expect the thermal disk luminosity to be similar to Eddington at most viewing angles. The timescale for accretion to exceed the Eddington limit was found to be months to a year for WD tidal disruptions by \citet{MacLeod:2014gl}. Thus, during this Eddington limited phase, the implied disk luminosities are $L_{\rm d} \sim L_{\rm Edd} \approx 2.5 \times 10^{42} (\Mbh / 10^4 M_\odot)$ erg s$^{-1}$. And maximum temperatures are 
\beq\label{Tmax}
T_{\rm max} \sim 10^6 K  \left [ \left(\kappa \over 0.2 {\rm \ cm^2 \ g}^{-1} \right) \left(\eta \over 0.1  \right)  \left(\Mbh \over 10^4 M_\odot  \right) \right]^{-1/4} ,
\eeq  
where $\eta$ is the radiative efficiency of the disk $L=\eta \dot M c^2$ \citep{Miller:2015tj}, also see \citet{Haas:2012ci} for a similar estimate derived from a slim disk model. 
Further,  \citet{2015arXiv150704333D} have argued that deeply plunging tidal disruption events might preferentially form compact disks with characteristically high effective temperatures as a result of their larger relativistic precession of periapse. 
This temperature implies peak emission frequencies in the soft X-ray ($\nu_{\rm peak } \sim 6 \times 10^{16}$ Hz $\sim 0.25$ keV).  If the spectrum is characterized by a blackbody, the optical $\nu L_\nu$ is lower by a factor of $\sim 10^5$ than the peak, suggesting that thermal emission in the optical band is likely very weak. 

If an optically thick reprocessing layer  forms  from tidal debris that is scattered out of the orbital plane \citep{Loeb1997,Ulmer:1998vk,Bogdanovic:2004gd,Strubbe:2009ek,Guillochon:2013vh,Coughlin:2014fs}, the photosphere temperature might be  lower, bringing the peak of the spectral energy distribution into the ultraviolet. 
As noted by \citet{Miller:2015tj}, this is found to be the case in optically-discovered tidal disruption transients \citep{vanVelzen:2011gz,Gezari:2012fk,Chornock:2013dm,Arcavi:2014im,Holoien:2014go,Vinko:2015hv}.    \citet{Miller:2015tj} has also suggested that disk winds may lower the peak disk temperature by an order of magnitude by reducing the flux originating from the innermost radii of the disk.  If either a lower temperature photosphere enshrouds the disk \citep{Guillochon:2013vh} or strong disk winds lower the peak temperature, then the ultraviolet or even optical luminosity could conceivably reach brightness similar to the disk luminosity and MBH Eddington limit.

\citet{Clausen:2010vk} have shown that the irradiated unbound debris of a WD tidal disruption can  produce notable emission line spectral signatures when irradiated by photons from the accretion disk. Emission lines of carbon and oxygen are found to be particularly strong, with CIV at 1549 \AA\ able to persist at $\sim 10^{38}$ erg s$^{-1}$ for hundreds of days following the disruption. Similarly, [OIII] lines at 4363 and 5007 \AA\ should be quite strong $\sim 10^{36}-10^{37}$ erg s$^{-1}$ over timescales of hundreds of days. During the phase in which the thermonuclear transient is near peak these lines would be overwhelmed. But, as the transient fades, deep observations could look for the signatures of these emission lines which would be expected to slowly fade to the host galaxy level.

\subsection{Jet Production and Signatures}\label{sec:jet} 

Fallback and accretion rates realized by WD tidal disruptions regularly exceed the MBH's Eddington limit by a large margin. Thus, although thermal, accretion disk emission is limited to $\sim L_{\rm Edd}$, many of these systems should also launch relativistic jets \citep[e.g.][]{Strubbe:2009ek,Zauderer:2011ie,DeColle:2012bq}. These jets carry power which may depend on MBH spin and available magnetic flux as in the \citet{Blandford:1977uk} mechanism or they can even be powered solely by the collimation of the radiation field \citep{Sadowski:2015vr}. In either case, the isotropic equivalent jet power is expected to be a fraction of $\dot M c^2$ \citep[e.g. ][]{Krolik:2012da,DeColle:2012bq,Tchekhovskoy:2013gw,Sadowski:2015vr}.  As a result, for viewers aligned with the jet axis, the isotropic equivalent of the beamed emission from jetted transients can substantially outshine the thermal emission because $\dot M c^2 \gg \dot M_{\rm Edd} c^2$ \citep{MacLeod:2014gl}.
The jet power, the radiative efficiency, and the relativistic beaming factor along the line of sight are all thus uncertain to some degree \citep[See e.g.][section 3.1, for a description of these uncertainties in the context of tidal disruption jets]{Krolik:2012da}.

In the upper panel of Figure \ref{Fig:accretion} we map the accretion rate to a peak  beamed jet luminosity following a simple accretion rate scaling, $L_j=0.1 \dot M c^2$.  Under these assumptions, typical peak jet luminosities for events that lead to explosions are in the range of $\sim 10^{47} - 10^{50}$ erg s$^{-1}$  with rise timescales of $\sim 10^{2}-10^4$ seconds.  These powerful jets will result in short-duration, luminous, flares with the bulk of the jet's isotropic equivalent luminosity Doppler boosted to emerge in high energy bands \citep{MacLeod:2014gl}. 
At viewing angles along the jet beam, the observer would see high energy emission arising either from Compton-upscattering of the disk's photon field (which peaks in the UV/soft X-ray, see equation \ref{Tmax}) or  by internal dissipation within the jet \citep{vanVelzen:2011gu,vanVelzen:2013cs,Krolik:2012da,DeColle:2012bq,Tchekhovskoy:2013gw,MacLeod:2014gl}.  In the cases of transients thought to be relativistically beamed tidal disruption events (Swift J1644+57, J2058+05, and J1112-8238) the spectrum can be explained by either alternative \citep{Bloom:2011er,Burrows:2011kz,Cenko:2012if,Pasham:2015ul,2015arXiv150703582B}.  The observed luminosity of these transients suggest that X-ray efficiencies ($L_X = \eta_X L_j$) of order unity may be typical \citep{Metzger:2012hr}. Further, the jet itself can be relatively long-lived, with accretion exceeding the MBH's Eddington limit for timescales of months to a year  \citep[e.g. Figure 2 of][]{MacLeod:2014gl}. 

 If the \citet{Blandford:1977uk} mechanism supplies the jet power, there may be a strong jet power (and thus luminosity) dependence on the energy source -- the MBH's spin and the magnetic field in the disk midplane \citep[e.g.][]{Tchekhovskoy:2013gw}. Our illustrative assumption of 10\% efficiency is  applicable for nearly maximally rotating MBHs. \citet{2005ApJ...630L...5M} show, through numerical simulations, a steep dependence of jet power on spin, where $\sim$7\% of $\dot M c^2$ is ejected in jet power for Kerr spin parameter of $a/M=1$, while for $a/M=0.9$, efficiencies of 1\% result. Thus, if the typical MBH in this mass range has a low spin parameter, we might expect systematically less luminous jets than if MBHs are universally rapidly-spinning.

Both on and off of the beaming axis, radio emission generated from interaction between the relativistic outflow and the external medium should trace the jet activity. Along the jet axis, the fundamental plane of BH activity \citep{Merloni:2003kv} relates X-ray and 5GHz Radio luminosity to BH mass. Ignoring uncertainties, $\log L_R  = 6.32 + 0.82 \log \Mbh + 0.62 \log L_X$ \citep{Merloni:2003kv,Miller:2011jq}.   Given a flare accretion rate, we can also use the fundamental plane to estimate characteristic radio luminosities. For a boosted luminosity of $L_X = 10^{48}$ erg s$^{-1}$ we might infer an intrinsic luminosity of $L_X = 10^{46}$ erg s$^{-1}$ given $\Gamma_j = 10$ \citep[e.g.][]{Metzger:2012hr,DeColle:2012bq,Mimica:2015jw}. Then, with $\log \Mbh = 4$, the predicted 5GHz radio luminosity would be $L_R \sim 10^{38}$ erg s$^{-1}$.  This characteristic radio luminosity may be de-boosted to off-axis viewing angles with the transformation $ L_\nu = L_{\nu,0} (1-\beta_j \cos(i_{\rm obs}) )^{\alpha_{\rm s}-2}$, where $L_\nu$ is  the observed and  $L_{\nu,0}$ is  source luminosity density at frequency $\nu$. The jet velocity is $\beta_j = v_j/c$ and $\cos(i_{\rm obs})$ accounts for the observer's viewing angle. The spectral index is $\alpha_{\rm s}$.  This expression suggests a difference of a factor of $\sim 40$ for on vs off axis viewing angles ($\cos(i_{\rm obs})$ of 0 or 1) with a jet Lorentz factor $\Gamma_j = 10$ and $\alpha_{\rm s} = 1.3$ \citep{Zauderer:2011ie,Miller:2011jq}. With $\alpha_{\rm s} =1$, a factor of $200$ can easily arise.  

To illustrate the time dependent properties of this radio emission, we compare to afterglows of decelerating jets calculated by \citep{vanEerten:2010gw}. The kinetic energy of these jets is $E_K = 2\times10^{51}$ erg, which implies a fraction of the rest energy $\eta_j = E_K / M_{\rm acc} c^2 \approx 6 \times 10^{-3}$ where the accreted mass $M_{\rm acc} = 0.2 M_\odot$ goes into powering the jet. The jet opening angle is assumed to be $\theta_j = 0.2$. The isotropic equivalent power is $\approx 10^{53}$ erg, implying a beaming fraction of $\approx 0.02$ and a mean Lorentz factor of $\Gamma_j \sim 7$. The surrounding medium number density is assumed to be $n=1$ cm$^{-3}$. We might expect that the jet-medium interaction dynamics would proceed somewhat differently in the case of a tidal disruption event, where the jet is not an impulsively powered blastwave. However, the jet power is dominated by the time of peak accretion, which comes long before the afterglow peaks at longer wavelengths. Thus, the impulsively powered blastwave provides a reasonable zeroth-order approximation of the afterglow properties.

\begin{figure}[tbp]
\centering
\includegraphics[width=0.49\textwidth]{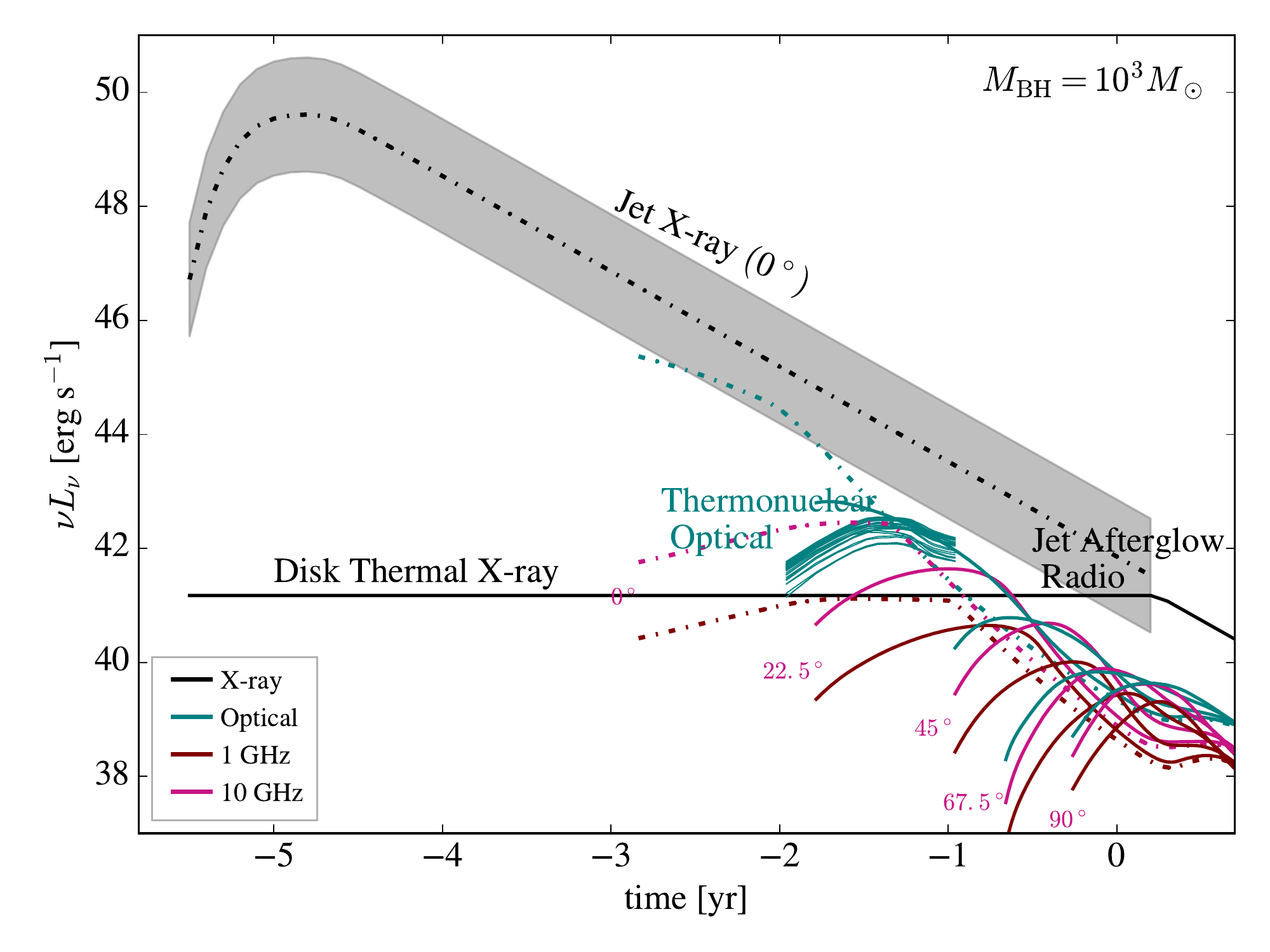}
\caption{This diagram depicts the observed multiband lightcurve of a hypothetical deep-passing WD tidal disruption by a $10^3 M_\odot$ MBH.  We have assumed that a jet is launched carrying kinetic energy of $6\times10^{-3}$ of the accreted rest energy  of $0.2 M_\odot$ from the disruption of a $0.6 M_\odot$ WD, or $2 \times 10^{51}$ erg. Line styles denote different viewing angles with respect to this jet axis. Dot-dashed lines are along the jet axis, while solid lines denote an off-axis perspective. Colors show different wavelengths of emission from X-ray to 1 GHz radio. For a viewer along the jet axis, a luminous jet may precede an optical-wavelength thermonuclear transient. The thermonuclear transient significantly outshines both jet afterglow and disk thermal emission at optical wavelengths. For off-axis events detected in the optical, accompanying X-ray emission from the accretion disk along with a radio afterglow complete the multiwavelength picture, and significantly distinguish these thermonuclear transients from other SNe.  }
\label{Fig:summary}
\end{figure}

Our results are visualized in Figure \ref{Fig:summary}, which represents the multi-wavelength and multi-viewing angle perspective of a deep-passing WD tidal disruption event.  Properties that can be observed only from along the jet axis (here $0^\circ$), are drawn with dot-dashed lines. Properties visible to off-axis observers are shown in solid lines. Colors denote the wavelength of the emission plotted. We show X-ray (0.2 -10 keV), Optical (V-band) as well as 1 and 10 GHz radio emission. Efficiencies in these bands for the disk and jet components are computed assuming that the disk spectrum is approximately a blackbody at $T_{\rm max}$, equation \eqref{Tmax}, and the jet spectrum is boosted by $\Gamma_j^2$ in frequency and apparent luminosity.  For an observer at $0^\circ$, a luminous X-ray jet proceeds an optical afterglow, which is followed by the thermonuclear transient. The radio afterglow follows as the thermonuclear transient begins to fade in the optical. The X-ray jet is depicted with a shaded region to represent the high level of variability observed in Swift J1644+57, J2058+05, and J1112-8238 \citep[e.g.][]{Saxton:2012ip,Pasham:2015ul,2015arXiv150703582B}. For an off axis observer, early time emission is dominated by the accretion disk, whose peak is in the soft X-ray. Disk optical and UV are not plotted here because they are $10^5$ and $10^4$ times less luminous, respectively (assuming blackbody emission from the disk). At off-axis angles, the thermonuclear transient is substantially brighter (and shorter duration) than the optical afterglow. The radio afterglow follows and peaks with timescales of 0.1 - 1 year, depending on viewing angle.

\section{Event Rates and Detectability of WD Tidal Disruption Signatures}\label{sec:detect}

In this section, we use our estimates of the observable properties of WD tidal disruption events to study the rate and properties of events detectable in current and upcoming high energy monitors and optical surveys. 

\subsection{Specific Event Rate}
Following the calculation presented in \citet{MacLeod:2014gl}, we estimate that the specific (per MBH) event rate of WD tidal disruptions is of order $\dot N_{\rm MBH} \sim 10^{-6} {\rm \ yr}^{-1}$.  This estimate is based on the method of \citet{Magorrian:1999fd}. In so doing, we assume that the MBH mass-velocity dispersion relation can be extrapolated to MBH masses of $\sim 10^3 - 10^5 M_\odot$. This extrapolation suggests that the MBH is energetically dominant within a region of radius
$r_{\rm h} = G M_{\rm bh} / \sigma_{\rm h}^2   = 0.43 \left( M_{\rm bh} / 10^5 M_\odot \right)^{0.54} \ {\rm pc} $
where $\sigma_{\rm h}$ is the velocity dispersion, adopted from the  $\Mbh-\sigma$ relation, $\sigma_{\rm h} = 2.3 \times 10^5 \left( \Mbh / M_\odot \right)^{1/4.38} \text{ cm s}^{-1}$  \citep{Kormendy:2013vg}.   
We assume that the stars within this radius are distributed in a steep cusp with $\nu_* \propto r^{-3/2}$, as representative of a relaxed stellar distribution \citep{Frank:1976tg,Bahcall:1976kk,Bahcall:1977ea}. 
We normalize the density of stars in the central cluster (at radii less than $r_{\rm h}$ from the MBH) by assuming that the enclosed stellar mass is equal to the MBH mass \citep{Frank:1976tg}.  
While these assumptions are motivated by observed distributions of stars around supermassive BHs \citep[e.g.][]{Lauer:1995dm,Faber:1997fn,Magorrian:1998cs,Syer:1999gp,Kormendy:2013vg}, there is significant uncertainty in their extrapolation to low mass. 
Finally, we assume that the fraction of stars that are WDs is $f_{\rm WD}=0.1$, and that they are distributed in radius proportionately with the rest of the stellar distribution. 
For more discussion of the derivation of this rate we refer the reader to \citet{MacLeod:2014gl}. 

Another possible host system for lower or intermediate mass MBHs is globular clusters. The tidal disruption event rate (of main sequence stars) in these clusters is relatively low, $\sim 10^{-7}{\rm \ yr}^{-1}$ \citep{RamirezRuiz:2009gw}. A smaller fraction still of these events would be WD tidal disruptions. If 1\% of the events are WD disruptions, the event rate per globular cluster MBH would be $ \sim 1 {\rm \ Gyr}^{-1}$ \citep[e.g.][]{Baumgardt:2004jf,Baumgardt:2004dx,Haas:2012ci,Shcherbakov:2013hf,Sell:2015wh}. 
 Each galaxy hosts many globular cluster systems, but not all of them will necessarily host a MBH. By contrast, evidence from the Milky Way's globular cluster system is that most globular clusters do not host MBHs at the present epoch \citep{Strader:2012tw}. If, for example, one in 100 clusters hosted a MBH, then this would mitigate the potential enhancement of a given galaxy hosting up to $\sim100$ globular clusters. Thus, with the evidence present, we suggest that MBHs in dwarf galaxies may be the primary source of WD tidal disruption events. 

Only a fraction of disruption events pass close enough to the MBH to lead to a thermonuclear transient. We can estimate the fraction of disruption events that lead to a thermonuclear transient based on the relative impact parameters needed. We denote critical impact parameter leading to thermonuclear ignition as $\beta_{\rm thermo}$.  When the phase space of the loss cone is full because it is repopulated efficiently in a orbital period, the fraction of tidal disruption events with $\beta>\beta_{\rm thermo}$ is $(\beta_{\rm thermo} / \beta_{\rm ml} )^{-1}$, where $\beta_{\rm ml}$ is the impact parameter at which the WD would start lose mass. This is expected to be the case for WDs in clusters around MBHs \citep[See Figure 4 of ][]{MacLeod:2014gl}.  For WDs, $\beta_{\rm ml}\approx0.5$ \citep{Guillochon:2013jj,MacLeod:2014gl} and $\beta_{\rm thermo} \approx3$ \citep{Rosswog:2009gg}, so we can expect a fraction $f_{\rm thermo}\approx1/6$ of events to result in a thermonuclear transient. 
When the MBH mass becomes too large $\gtrsim 10^5 M_\odot$, the deeply plunging orbital trajectories will lead to the WD being swallowed by the MBH with no possibility to produce a flare or thermonuclear transient \citep{MacLeod:2014gl}.

\subsection{The MBH Mass Function: Estimating the Volumetric Event Rate}\label{sec:mf}
To convert this specific event rate to a volumetric rate, we must then consider the space density of MBHs hosting dense stellar clusters. 
 \citet{Sijacki:2014vg} find a MBH number density per unit MBH mass of $\Phi(\Mbh) \approx 10^{7}$~Gpc$^{-3}$~dex$^{-1}$ for $\Mbh \sim 10^6$ and redshifts $z\lesssim2$ in the Illustris simulation. In the following, we illustrate the range of possibilities using three examples for the extrapolation of this mass function present themselves.
 \begin{enumerate}
 \item  One possibility is that the mass function for $\Mbh < 10^6 M_\odot$ is flat, and  $\Phi(\Mbh) = 10^7$~Gpc$^{-3}$~dex$^{-1}$. 
 \item A second possibility is that the slope of approximately $\Phi(\Mbh) \propto \Mbh^{-1/2}$ observed in the MBH mass function for MBHs $10^7 M_\odot < \Mbh < 10^9 M_\odot$ (ie $10^7 M_\odot$ MBHs are approximately $10\times$ more common than $10^9 M_\odot$ MBHs) extends to masses below  $10^6 M_\odot$. In this case we normalize the distribution to $\Phi(10^7 M_\odot) \approx 5\times 10^{6}$~Gpc$^{-3}$~dex$^{-1}$.  If the occupation fraction of MBHs in dwarf galaxies is of order unity, it would indicate a rising MBH mass function to lower masses \citep[e.g. work by][constrains the density of dwarfs in the local volume]{Blanton:2005gz}.
 \item A third possibility is that no MBHs below $\sim 10^6 M_\odot$ exist, although new detections of MBHs with masses $\sim 10^4 M_\odot$ make this option appear unlikely \citep{2009Natur.460...73F,2013ApJ...775..116R,2015arXiv150607531B}. 
 \end{enumerate}
 Each of these mass functions can be integrated over MBH mass to give the volume density of MBHs in the that disrupt WDs, $n_{\rm MBH}$. We adopt limits of $10^3 - 10^5 M_\odot$ here, since lower-mass MBHs are less likely to host tightly-bound stellar clusters, and higher mass MBHs often swallow WDs whole rather than disrupting them \citep[e.g.][]{MacLeod:2014gl}. 

Adopting the flat extrapolation of the MBH mass function to lower masses implies $n_{\rm MBH} \approx 2 \times 10^7$~Gpc$^{-3}$. If we instead assume that the mass function continues to rise to lower MBH masses, the volume density of $10^3 - 10^5 M_\odot$ is $n_{\rm MBH} \approx 4 \times 10^8$~Gpc$^{-3}$. 
The volumetric event rate of thermonuclear transients accompanying WD disruptions in dwarf galaxies can then be estimated as
\beq\label{volrate}
\dot N_{\rm vol} \approx 1.7 \left( \frac{\dot N_{\rm MBH}}{10^{-6} {\rm yr}^{-1}} \right) \left( \frac{n_{\rm MBH}}{10^7 {\rm Gpc}^{-3}} \right)   \left( \frac{f_{\rm thermo}}{1/6} \right){\rm yr^{-1} \ Gpc}^{-3}.
\eeq
where a factor of 2 higher gives the appropriate scaling for a flat extrapolation of the MBH mass function, and a factor of 40 gives the appropriate scaling for a MBH mass function that rises to lower masses. We will use these examples to illustrate most of the remainder of our analysis. In Section \ref{sec:disc_bhmf}, we allow for a MBH mass function with free power-law index below $10^7 M_\odot$.

\subsection{Detecting Thermonuclear Transients with LSST}

\begin{figure*}
\centering
\includegraphics[width=0.99\textwidth]{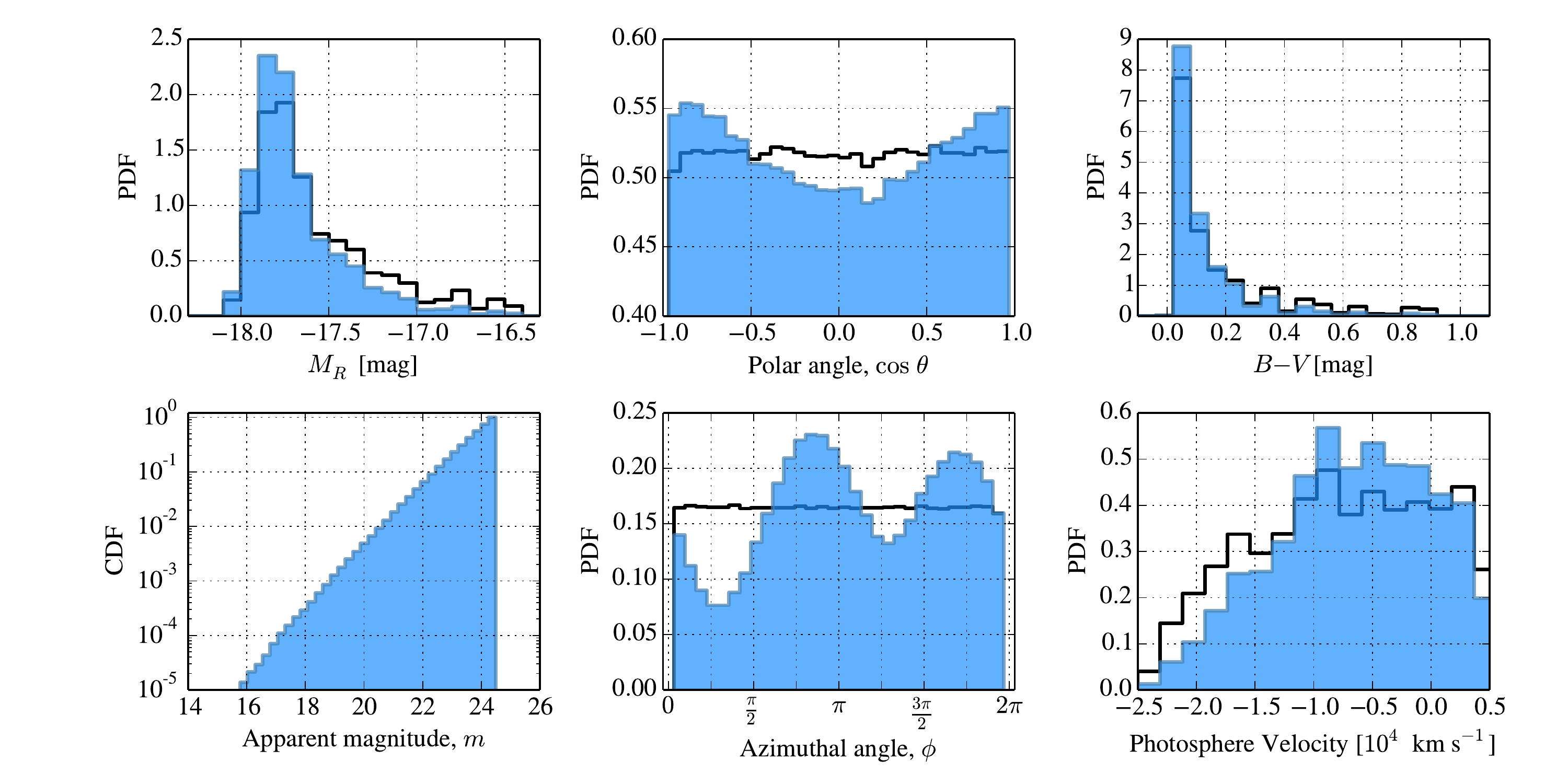}
\caption{A Monte Carlo realization of thermonuclear transients captured by a magnitude limited transient optical survey. For this figure we adopt a limiting magnitude of 24.5, the R-band single exposure limit for LSST. Black lines show the intrinsic distributions of event properties, including absolute magnitude, viewing angles, color and photosphere velocity evaluated near R-band peak at $t=16$ days. Blue histograms show how these intrinsic properties are mapped to the detected events. Detectible events exhibit a preference for particular viewing angles, and the detected absolute magnitude distribution is centered around the brighter events. The photosphere velocity is assessed by the doppler offset of the minimum of the SiII 6355~\AA\ line and scales as $(\Mbh/M_{\rm WD})^{1/6}$. The event rate of detection given these survey properties and a flat extrapolation of the MBH mass function to masses below $10^6 M_\odot$ is $\approx 14$ yr$^{-1}$. }
\label{Fig:MC}
\end{figure*}

With an estimate of the volumetric event rate 
as guidance, we can now estimate the detection rate and detectable distributions of events in upcoming optical surveys. We will focus on the Large Synoptic Survey Telescope (LSST), which, when operational, will survey south of +10 degrees declination and have an approximately 3 day cadence, making it well suited to catching these relatively rapid transients \citep{Collaboration:2009vz}. The anticipated R-band limit for a single 15 second LSST exposure is 24.5 magnitude \citep{Collaboration:2009vz}.  

We calculate that the brightest R-band viewing angles ($M_R\approx -18$) could be detected by LSST out to a luminosity distance of $2.0$ Gpc (redshift, $z=0.37$), if we adopt a typical R-band extinction, $A_R$, of 1 magnitude \citep[we assume a WMAP9 Cosmology,][]{Hinshaw:2013dd}.   This extinction implies $A_V \sim 1.25$  \citep[e.g.][]{Hendricks:2012cw}. 
This is higher than most observed type Ia SNe discovered \citep[][]{Holwerda:2008jv},
 but we expect that some lines of sight into galactic nuclear regions would be heavily extincted \citep{Strubbe:2009ek}. 
However, the thermal tidal disruption flare discovered by \citet{Gezari:2012fk} exhibits relatively low reddening, consistent with $A_V\approx0.25$ for $R_V = 3.1$.
Given this maximum viewing distance we use equation \eqref{volrate} to infer an event rate of $ \approx 44$ events per year within the 13.3 Gpc$^3$ volume inclosed by $z=0.37$ for MBH mass function option (1) in Section \ref{sec:mf}, a flat extrapolation to lower MBH masses. We find $\approx 890$ events per year for MBH mass function (2), which rises to lower MBH masses.

There is, however, substantial variation with viewing angle seen in the lightcurves. We therefore perform a Monte Carlo sampling of events in viewing angle and redshift with viewing angles sampled isotropically and redshifts sampled according to the comoving volume. In this simple model, we assume all events have the lightcurves of the representative tidal disruption event described in this paper.  
In our Monte Carlo simulations, we find an LSST-detectable event rate of 
\beq\label{rateLSST}
\dot N_{\rm LSST} \approx 14\; (290) \; {\rm yr}^{-1} 
\eeq
for MBH mass function options (1) and (2), respectively. 
This rate scales with the same physical parameters as the volumetric rate of equation \eqref{volrate}, and with the adopted extinction as $10^{-3 A_R / 5}$ (ie, with no extinction, the detectable event rate would be a factor of $\sim 4$ higher than with $A_R = 1$).  This calculation suggests that this distribution of events in distance and in peak brightness implies that $\sim 60$\% of events are detected within the maximum volume based on the observed sky area, a conclusion that also holds for other survey limiting magnitudes. 

In Figure \ref{Fig:MC}, we compare intrinsic distributions of events to those that could be observed with LSST.  In black, we plot the intrinsic distributions of source properties, while the shaded blue histograms show the distributions of detected events.  The left two panels look at source absolute and apparent magnitude distributions.  The detectable distribution is biased toward the viewing angles which generate the brightest transients. This can be seen both in the absolute magnitude distribution and in the center panels which show the viewing angle distributions.  The two right-hand panels quantify some of the ways that this viewing angle preference propagates into source properties evaluated at $t=16$ days, near R-band peak. We find that detected events are slightly biased toward lower $B-V$ (bluer) colors at day 16.  Lower photosphere velocities are also preferred, with those blue-shifted by more than 15,000 \kms\ somewhat less likely to be observed.  In summary, though, these selection effects are mild, and do not suggest than one particular viewing angle, color, or photospheric velocity is strongly preferred.

\subsection{Detecting Beamed Emission with High-Energy Monitors}

In Section \ref{sec:jet} we outlined a case for the production of jets and beamed emission in tidal disruptions of WDs. Here, we examine the detection of this jetted emission by high-energy monitors like the Swift Burst Alert Telescope (BAT) \citep{2013ApJS..209...14K}. 
The BAT is sensitive in the hard X-ray channel $(\sim 15 - 150{\rm ~keV})$ and is thus well suited to detection beamed emission generated by a tidal disruption-fed accretion flow. 
We can use the BAT threshold $\sim 2\times10^{-10} (t/20 {\rm ks})^{-1/2}$~erg~s$^{-1}$~cm$^{-2}$ (5 sigma) for an exposure time $t$.\footnote{Table 2.1 of BAT user's guide v6.3 available at {\tt http://swift.gsfc.nasa.gov/analysis}} This implies that a jetted transient for which the BAT-band luminosity is $10^{48}$~erg~s$^{-1}$ can be observed to a luminosity distance of $2 \times 10^{28}$~cm  (6.5 Gpc) or, with our assumed WMAP9 cosmology, a redshift of $0.96$. If the jet luminosity reaches $10^{49}$~erg~s$^{-1}$, then it may be detected to a redshift of 2.45.  

These redshifts imply that the $10^{48}$~erg~s$^{-1}$ and $10^{49}$~erg~s$^{-1}$  transients can be observed within cosmological volumes 11 and 65 times larger than the LSST volume for the thermonuclear transient.  Following from the volumetric event rate, we expect the BAT detection rate of WD disruption events that also produce a thermonuclear transient to be of the order of
\beq
\dot N_{\rm BAT} \approx 12 \ \rm{yr}^{-1},
\eeq
if we assume $10^{49}$~erg~s$^{-1}$ transients with a beaming fraction of $1/50$ $(\Gamma_{\rm j} \sim 7)$, that the BAT monitors 20\% of the sky \citep{2013ApJS..209...14K}, and a flat MBH mass function, option (1) in Section \ref{sec:mf}. If the typical transient is instead $\sim 10^{48}$~erg~s$^{-1}$, but other properties are similar, then the BAT rate  is $\approx 2 \ \rm{yr}^{-1}$ based on the smaller accessible volume.  
If the MBH mass function rises to lower masses, option (2) in Section \ref{sec:mf}, then we infer a detection rate of 
\beq
\dot N_{\rm BAT} \approx 230 \ \rm{yr}^{-1}
\eeq
for $10^{49}$~erg~s$^{-1}$  transients and $\approx 40 \ \rm{yr}^{-1}$ for $10^{48}$~erg~s$^{-1}$  transients.
The rates above are for transients that produce a thermonuclear transient. Less deeply-plunging tidal disruptions of WDs occur $f_{\rm thermo}^{-1} \approx 6$ times more frequently \citep{MacLeod:2014gl}. 

This calculation suggests that despite their substantially different peak luminosities at different frequencies, WD tidal disruption transients may be currently detectable by BAT at rates similar to what LSST will allow for in the future in the optical. Because high-energy emission precedes the thermonuclear transient, deep optical follow-up for  candidate disruption flares is highly desirable, and it offers the best present-day strategy for detecting these transients.  The thermonuclear transients detected via follow-up of beamed emission would lie along the jet-launching axis, perhaps lying perpendicular to the original orbital plane (although significant MBH spin could torque the debris stream or the inner accretion disk out of its original plane, changing the orientation of the jet/disk relative to the thermonuclear transient).

\section{Discussion}\label{sec:discussion}

\subsection{A Diversity of Thermonuclear Transients from WD Tidal Disruption}
In this work, we have examined in detail one model thermonuclear transient generated by the tidal compression and disruption of a carbon/oxygen WD. A primary caveat in extending the conclusions reached through examination of this model is that the tidal disruption process is likely to generate a diversity of transients.  However,  there is little reason to expect MBH-mass dependence on the generation of thermonuclear transients via tidal disruption. To linear order, the forces acting on the WD and timescale of passage can all be written in terms of the dimensionless impact parameter $\beta$.  Whether or not runaway burning takes place depends on the ratio of dynamical timescale to burning timescale. The passage timescale is not a function of MBH mass, because for $\beta=1$ encounters it is always equal to the WD dynamical timescale. Similarly, the bulk velocity of the unbound ejecta scales only weakly with MBH mass, $v_{\rm max} \propto  (\Mbh/M_{\rm WD})^{1/6}$. 
These simple scalings can be modified when encounters have pericenter distances similar to $r_{\rm s}$ and relativistic effects are important.  \citet{2015MNRAS.449..771G} found an increased effective impact parameter (stronger compression and mass loss than the Newtonian limit) for strongly relativistic encounters.

While strong dependence on MBH mass is not expected, the degree of burning should scale with both WD mass and impact parameter, $\beta$, as has been demonstrated by the simulations of \citet{Rosswog:2008gc} and \citet{Rosswog:2009gg}. At the extremes, these range from no burning in weakly-disruptive encounters to strong burning in deep encounters.
In particular, varying amounts of iron-group elements synthesized will affect the peak brightness of optical transients -- which are powered primarily through radioactive decay. 
Table 1 of \citet{Rosswog:2009gg} provides a summary of the nuclear energy release and iron-group synthesis in their runs.  The mass of iron group elements in explosive events ranged between $\sim0.01-0.7 M_\odot$. In cases where burning is incomplete, we might expect partial burning in some portions of the WD to manifest itself as large amounts of incompletely burned $\alpha$-chain elements in the spectrum.  
The iron-group mass is largest in deeper encounters, and in those involving massive white dwarfs. For example, a $0.2M_\odot$ WD in a $\beta=12$ encounter produces $0.034 M_\odot$ of iron-group material, but a $1.2 M_\odot$ WD in a $\beta=1.5$ orbit produces a similar quantity. \citet{Rosswog:2009gg} also find the degree of iron-group synthesis to be a strong function of $\beta$; a $1.2 M_\odot$ WD in a $\beta=2.6$ orbit produces $0.66M_\odot$ of iron group elements.  
This range of iron-group masses produces a spread in lightcurve peak brightnesses (and varying WD mass likely to a range of peak timescales, with lower masses resulting in more rapid peak) across the range of possible disruptions. Interestingly, however, the viewing-angle diversity explored in this paper is of a similar magnitude, and may encompass some of the diversity in possible WD-MBH combinations.

The simulation presented in this paper used a C/O WD, but many lower-mass WDs are composed primarily of helium. Helium is significantly easier to burn than carbon and oxygen, with detonations being possible at lower temperature and densities \citep{2009ApJ...696..515S,2013ApJ...771...14H}, and thus perhaps at more grazing impact parameters, decreasing $\beta_{\rm thermo}$ and increasing $f_{\rm thermo}$. Helium WDs are also significantly lower in density than C/O WDs, enabling their disruption by higher-mass MBHs before they are swallowed whole, with disruptions by $M_{\rm bh} \gtrsim 10^{6} M_{\odot}$ being possible. \citet{2013ApJ...771...14H} show that incomplete burning with large mass fractions of $^{40}$Ca, $^{44}$Ti, $^{48}$Cr are a common outcome of He detonations in conditions typical of WD - MBH encounters \citep{Rosswog:2008gc,Rosswog:2009gg}. Thus, such disruptions were recently proposed by \citet{Sell:2015wh} to explain a calcium-rich gap transients \citep{2012ApJ...755..161K}, faint Ia-like events with large velocities, fast evolution, and occurring preferentially in the outskirts of giant galaxies where unseen dwarf hosts (which would potentially harbor moderate-mass MBHs) may lie.  \citet{2015MNRAS.452.2463F} emphasizes that kinematic evidence may be key in disentangling the origin of calcium-rich transients, and argues that those discovered so far have line-of-sight velocities suggestive of being ejected from disturbed galactic nuclei -- but only exploding much later. Although we predict strong line-of-sight velocities, at face value \citet{2015MNRAS.452.2463F}'s position-dependent kinematic evidence is not consistent with a prompt optical transient following an encounter with a MBH.

 Although a systematic study is beyond the scope of the present paper, we intend to explore a range of encounters and the diversity of possible optical thermonuclear transients in future work.

\subsection{Are ULGRBs WD Tidal Disruptions?}
\citet{Shcherbakov:2013hf} offer the tidal disruption of a WD as an explanation for the underluminous, long GRB 060218 and its accompanying SN 2006aj. While our calculations show that the properties of the  high-velocity  SN 2006aj is not consistent with  the tidal disruption of a WD, this suggestion  put forward the intriguing possibility that high energy flares from WD tidal disruption may already be lurking in existing data sets. \citet{Levan:2014iz} and \citet{MacLeod:2014gl} followed up on this suggestion but considered instead 
the emerging class of ultralong gamma ray burst (ULGRB) sources which has been reviewed in detail by \citet{Levan:2014iz} and \citet{2015arXiv150603960L}. At present, it remains uncertain whether these events form a distinct population of high-energy transients or the tail of the long gamma ray burst duration distribution \citep{2015arXiv150603960L}.  If these objects do, in fact, represent a distinct class of transients, several possible scenarios present themselves to explain their emission. Suggested progenitor models include collapsars, the collapse of giant or supergiant stars, and beamed tidal disruption flares, particularly those from WD disruption \citep{Levan:2014iz,MacLeod:2014gl,2015arXiv150603960L}. In addition to their long duration, the ULGRBs are highly variable in a manner reminiscent of the relativistically beamed tidal disruption flares Swift J1644+57 and Swift J2058+05 \citep[e.g.][]{DeColle:2012bq,Saxton:2012ip,Pasham:2015ul}. 

Of particular relevance to our study are searches for accompanying SN emission in the optical and infrared \citep[which are reviewed in detail by][in their Section 3.6]{2015arXiv150603960L}.  In the case of GRB 130925A at $z=0.35$, the afterglow appeared as highly extinguished and no optical afterglow emission or SN signatures were detected. SN emission from GRB 121027A at $z\approx 1.7$, would be challenging to detect.  In the cases of GRB 101225A and 111209A red excess is observed by \citet{Levan:2014iz} in the afterglow's lightcurve evolution. Figure 6 of \citet{Levan:2014iz}  compares this to a SN 1998 bw template. GRB 101225A shows a mild reddening only about a half magnitude brighter than the 1998 bw template and a  significantly bluer SED than the 1998 bw template at of order 10 days after the outburst \citep[Figure 7 of ][]{Levan:2014iz}. 

The lightcurve of the SN accompanying GRB 111209A was recently  published by \citet{2015Natur.523..189G}. The SN is extremely bright, $M_{\rm bol}\sim -20$, which rules out a thermonuclear transient from WD tidal disruption on the basis of the required nickel mass in \citet{2015Natur.523..189G}'s best fit for a thermonuclear power source, $\approx 1 M_\odot$ in $3 M_\odot$ of ejecta. This transient is $\sim 3 \times$ brighter than even the Ic's typically associated with GRBs, leading \citet{2015Natur.523..189G} to suggest a magnetar central engine model. It is also an order of magnitude brighter than the brightest WD tidal disruption transients studied here, suggesting that a tidal disruption is unlikely to be responsible for GRB 111209A. In theory, the explosion of a stripped core of a more-massive star could explain the energetics of the event \citep{Gezari:2012fk,2013ApJ...777..133M,2014ApJ...788...99B}, but the occurrence rate of deep encounters of tidally stripped stars is unknown. 

Optical and infrared SN searches for future  transients in the ULGRB category should offer further valuable constraints on whether WD tidal disruption is consistent with their origin. In particular, we now offer better description of the expected tidal thermonuclear lightcurves and spectra. We expect only a fraction $f_{\rm thermo} \approx 1/6$ to be accompanied by any SN-like transient.
Even a sample of $\sim 10$ events with follow-up observations could be used to provide strong evidence for or against a a WD tidal disruption origin based on the fraction accompanied by thermonuclear transients.

\subsection{Strategies for Identifying WD Tidal Disruptions}
In this paper we have identified the photometric and spectroscopic properties that distinguish WD tidal disruption transients from more-common SNe. Even so, it is worthwhile to compare the volumetric rate of thermonuclear transients generated by tidal disruptions to the most commonly observed thermonuclear explosions of WDs, type Ia SNe. Ia SNe are far more common, with event rates $\approx 5 \times 10^4 {\rm \ yr^{-1} \ Gpc}^{-3}$ \citep[e.g.][]{Graur:2014jp} as compared to $\sim {\rm few} {\rm \ yr^{-1} \ Gpc}^{-3}$.This comparison makes apparent the degree of challenge that will be faced by next generation surveys in identifying and recovering these and other exotic transients from amongst a vast quantity of SNe. 

Rather than relying on detection of only the thermonuclear transient, we suggest that the multi-wavelength signatures of WD tidal disruptions is what makes these transients truly unique. These signatures are generated, as described in Section \ref{sec:accretion} and Figure \ref{Fig:summary}, through a combination of nuclear burning and accretion power. Further, our examination of the relative sensitivity and luminosities of present high-energy and next-generation optical instrumentation and transients suggests that survey efforts may uncover WD tidal disruptions at similar rates in these two wavelengths. 

The best present-day strategy to  successfully uncover and firmly identify WD tidal disruption transients is to search for beamed emission with high energy monitors like Swift's BAT. Optical follow-up is a critical component of this strategy and could be used to constrain or detect the presence of a thermonuclear transient. If each high-energy-detected event were a tidal disruption, we would expect $f_{\rm thermo} \approx 1/6$ to be accompanied by a thermonuclear transient. 

When LSST comes online, a second potential strategy will emerge. This could 
 involve searching for optical transients with photometric and spectroscopic properties similar to those described here, and following-up viable candidate events at X-ray and radio wavelengths for accretion signatures. For example, the thermal emission from the accretion disk (eg. $L\sim 10^{42}$~erg~s$^{-1}$) would be visible in a 10~ks XMM-Newton exposure at $z\lesssim0.3$ \citep{2001A&A...365L..51W}, offering a valuable constraint on the MBH accretion that accompanies the transient. 
If the accompanying accretion flow launches a jet, we expect a radio afterglow to follow the optical transient, as described in Section \ref{sec:accretion}.

\subsection{Uncovering the Mass Distribution of Low-Mass MBHs}\label{sec:disc_bhmf}

This paper has examined the multiwavelength characteristics of transients that arise from WD-MBH tidal interactions. 
We find that the transients accompanying these interactions should be luminous and detectable in the optical and at high energies to redshifts of $z\sim 0.35$ (LSST, thermonuclear transients) or $z\gtrsim1$ (BAT, beamed emission from accretion flow).  We showed in Section \ref{sec:mf} that the largest uncertainty in estimating the detection rate is the uncertainty in the MBH mass function's extrapolation from well-known MBHs with masses of $10^6 - 10^9 M_\odot$ down to MBHs with masses of $\sim 10^3 - 10^5 M_\odot$.  We computed event rates for two possibilities, these are that we either extrapolate the value, or the slope of the MBH mass function, $\Phi (\Mbh)$, to lower masses.

If we extrapolate the slope of the MBH mass function to lower masses, a large number density of MBHs in the intermediate mass range will exist in the volume probed by WD disruption transients. Under this assumption WD disruption transients would be discovered at rates up to hundreds per year both by LSST and by high energy wide-field monitors like BAT. If instead the mass function remains relatively flat below $10^6 M_\odot$, we should detect tens of events per year. Finally, if no intermediate-mass MBHs exist, we should expect very few WD tidal disruption events and their associated signatures. Those that occurred would arise from rare encounters between maximally spinning MBHs with mass $\sim 10^6$ in which the orbital plane is aligned with the spin plane \citep{2012PhRvD..85b4037K}. Deep encounters, those than can produce thermonuclear transients, would be rarer still as they require an impact parameter well inside the tidal radius $(\beta \gtrsim 3)$ and thus require even more finely tuned conditions of MBH spin and orbital orientation.

\begin{figure}[tbp]
\begin{center}
\includegraphics[width=0.49\textwidth]{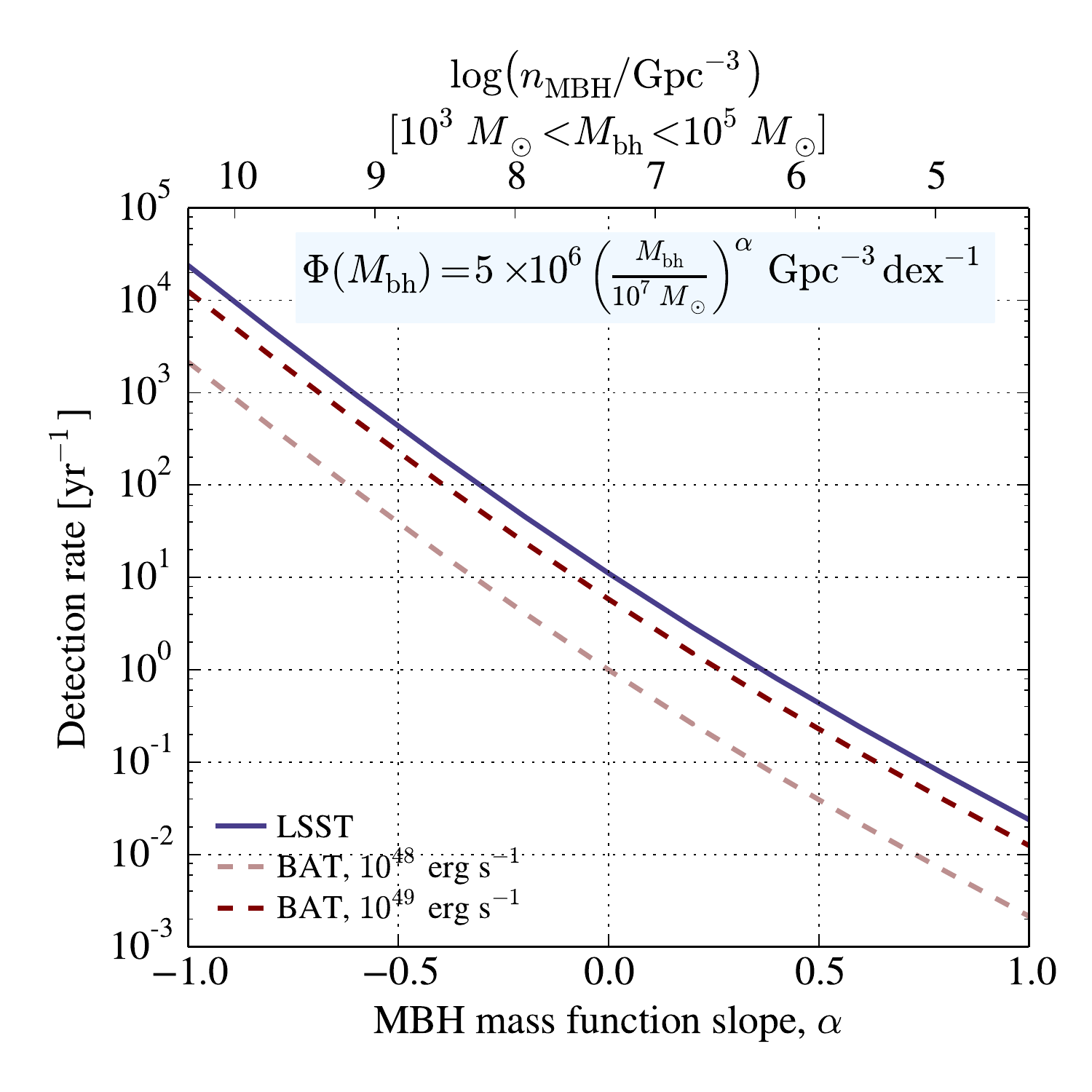}
\caption{ Effect of the slope of the MBH mass function below $10^7 M_\odot$ on the detection rate of WD tidal disruption transients.  This diagram assumes a simple, power law MBH mass function, $\Phi(\Mbh) \propto \Mbh^{\alpha}$, shown in the figure.  The power law slope of the low-mass MBH mass function determines the number density of MBHs with masses of $10^3-10^5 M_\odot$, whose density, $n_{\rm MBH}$, is shown on the upper x-axis. With detection rates ranging from tens to hundreds per year for flat to rising mass functions, it should be possible to place meaningful constraints on the allowed extrapolation of the MBH mass function to low masses within several years. The BAT transients plotted are only those deep passages that generate an accompanying thermonuclear transient (a fraction $f_{\rm thermo} \sim 1/6$ of the total events). The more common non-thermonuclear disruptions may also produce jets and high-energy transients, as described in Section \ref{sec:accretion}. 
\label{Fig:mf}}
\end{center}
\end{figure}

In Figure \ref{Fig:mf}, we illustrate the influence of the low-mass MBH mass function on the detection rate of WD tidal disruption transients associated with deeply-plunging WD tidal disruptions. We assume a simple power-law mass function in which $\Phi(\Mbh) \propto \Mbh^{\alpha}$, approximately normalized to the volume density of $10^7 M_\odot$ MBHs in the local universe \citep[e.g.][]{Sijacki:2014vg}. 
The slope of the extrapolation of the MBH mass function, $\alpha$, has a dramatic effect on the number density $n_{\rm MBH}$ of MBHs with masses of $10^3 - 10^5 M_\odot$ (which we assume can ignite WDs in tidal encounters). We assume the volumetric event rate of equation \eqref{volrate}. 
 A remaining caveat is the unknown typical jetted-transient beamed luminosity, its dependence on MBH spin, and whether all super-Eddington tidal disruption accretion flows successfully launch jets \citep[see, e.g.][for further discussion]{Krolik:2012da,DeColle:2012bq,Tchekhovskoy:2013gw,vanVelzen:2013cs}. We marginalize over these uncertainties by showing two possible jet luminosities in Figure \ref{Fig:mf}, and we note that our formalism can be easily extrapolated to other assumptions.
The variation in volume density of MBHs implies dramatic differences in the detection rate of transients associated with deeply-passing WD tidal disruptions. These differences are sufficiently significant that a few years of monitoring with high energy monitors like BAT or in the optical with LSST should produce a catalog of transients that can be used to place meaningful constraints on the number density of low-mass MBHs in the universe.

\section{Summary and Conclusion}\label{sec:conclusion}

In this paper we have examined the properties of thermonuclear transients that are generated following the ignition of nuclear burning in a deep tidal encounter between a WD and a MBH. 
This burning produces iron group elements in the core of the unbound debris of the tidal disruption event, surrounded by intermediate mass elements and unburned material. As this debris expands, an optical-wavelength transient with appearance similar to an atypical type I SN emerges. The peak brightness and color of the transient's lightcurve are highly viewing angle dependent as a result of the asymmetric distribution of the expanding tidal debris. A strong spectral signature of these transients is P-Cygni lines strongly offset from their rest wavelength by the orbital motion of the unbound debris (see Figures \ref{Fig:anglespec} and \ref{Fig:veloffset}). 

These transients should be accompanied by accretion signatures driven by relatively prompt bound-debris stream self-interaction and accretion disk formation. Accretion signatures range from thermal emission of the accretion disk, expected to peak in the soft X-ray, to harder X-ray non-thermal beamed emission along a jet axis. Optical and radio afterglow emission may trace the launching of a jet at viewing angles away from the jet axis.  We analyze the relative detection rates of WD tidal disruption transients at high energy and optical frequencies given high-energy wide field monitors like Swift's BAT and LSST in the optical. Our results suggest that detection rates may be similar with these disparate survey strategies, and we suggest that the most constraining events may be those in which multiple counterparts of the disruption event are observed. In particular, the most promising present-day strategy is probably deep optical follow-up of high-energy flares of unusually long duration or variability, like the ULGRBs \citep{Levan:2014iz,2015arXiv150603960L}.

In closing, we note that the existence of these transients remains uncertain \citep[e.g.][]{Sell:2015wh}, just as the existence of the MBHs of intermediate masses $\sim 10^3-10^5 M_\odot$ remains uncertain \citep[e.g.][]{2013ApJ...775..116R,2015arXiv150607531B}. With detailed estimates of the potential properties of these transients,  either their detection or non-detection should be able to be used to place meaningful constraints on the prevalence of intermediate mass MBHs in the universe.

\begin{acknowledgements}
We thank Maria Drout for guidance and data used in Figure~\ref{fig:phillips}, and  Ryan Foley for helpful discussions and comments on an early version of this manuscript.  We are grateful to Laura Chomiuk and Sjoert van Velzen for comments on radio afterglows of tidal disruption events and to Sjoert van Velzen for sharing code to estimate their properties.  We further acknowledge helpful conversations with Roseanne Cheng, Lixin Dai, Julian Krolik, Tom Maccarone, Phillip Macias, Michela Mapelli, Cole Miller, Dheeraj Pasham, Martin Rees, Elena Rossi, Michele Trenti.  MM is grateful for the support of the Chancellor's Fellowship at UCSC. ER-R acknowledges financial support from the Packard Foundation, Radcliffe Institute for Advanced Study and  NASA ATP grant NNX14AH37G.  JG acknowledges support from Einstein Grant PF3-140108. 
SR was supported by the DFG under RO-3399/8-1, AOBJ-575415
and by the Swedish Research Council (VR) under grant 621-2012- 4870.
Part of the simulations have been performed on the facilities of
the The North-German Supercomputing Alliance (HLRN).

\end{acknowledgements}

\end{document}